\documentclass[10pt,twocolumn,letterpaper]{article}

\usepackage[pagenumbers]{cvpr} 

\usepackage{booktabs}
\usepackage{times}
\usepackage{amsmath,amssymb}
\usepackage{empheq}
\usepackage{graphicx}
\usepackage[font=normalsize]{caption}
\usepackage{subcaption}
\usepackage{floatrow}
\usepackage{enumitem}
\usepackage{xcolor}
\usepackage{multicol}
\usepackage{multirow}
\usepackage{tikz}
\usepackage{graphicx}
\usepackage{array}
\usepackage{colortbl}
\usepackage{tcolorbox}
\usepackage{makecell}
\usepackage{adjustbox}
\usepackage[accsupp]{axessibility}  

\usepackage{algorithm}
\usepackage{algpseudocode}

\usepackage{tikz}
\usetikzlibrary{spy}
\usetikzlibrary{arrows.meta}

\algnewcommand\algorithmicinput{\textbf{Input:}}
\algnewcommand\algorithmicoutput{\textbf{Output:}}
\algnewcommand\Input{\item[\algorithmicinput]}
\algnewcommand\Output{\item[\algorithmicoutput]}

\definecolor{cvprblue}{rgb}{0.21,0.49,0.74}
\usepackage[pagebackref,breaklinks,colorlinks,allcolors=cvprblue]{hyperref}

\def\paperID{14236} 
\def\confName{CVPR}
\def\confYear{2026}

\title{FlowSteer: Conditioning Flow Field for Consistent Image Restoration}

\author{Tharindu Wickremasinghe$^{1}$, Chenyang Qi$^{2}$, Harshana Weligampola$^{1}$,  Zhengzhong Tu$^{3}$, Stanley H. Chan$^{1}$ \\
$^1$Purdue University
$^2$HKUST
$^3$Texas A\&M University}

\newcommand{\RB}[1]{\colorbox{red!20}{#1}}
\newcommand{\BB}[1]{\colorbox{blue!20}{#1}}

\usepackage{booktabs,siunitx,multirow,colortbl}
\newcommand{\N}{\mathcal{N}}
\newcommand{\I}{\mathbf{I}}

\makeatletter
\newcommand{\AlgNote}[1]{%
  \Statex \hskip\ALG@thistlm \textcolor{gray!70}{\emph{#1}}%
}

\makeatother

\usepackage{tabularx}
\newcolumntype{Y}{>{\raggedright\arraybackslash}X} 

\usepackage{listings}
\usepackage[dvipsnames]{xcolor}
\definecolor{codebg}{rgb}{0.98,0.98,0.98}
\definecolor{codeframe}{gray}{0.8}
\definecolor{codekw}{rgb}{0.1,0.1,0.6}
\definecolor{codestring}{rgb}{0.6,0.1,0.1}
\definecolor{codecomment}{rgb}{0.0,0.5,0.0}

\lstdefinestyle{python-ddnm}{
  language=Python,
  basicstyle=\ttfamily\footnotesize,
  keywordstyle=\bfseries\color{codekw},
  stringstyle=\color{codestring},
  commentstyle=\itshape\color{codecomment},
  showstringspaces=false,
  numbers=left,
  numberstyle=\tiny\color{gray},
  stepnumber=1,
  numbersep=5pt,
  frame=single,
  rulecolor=\color{codeframe},
  backgroundcolor=\color{codebg},
  breaklines=true,
  tabsize=2,
  xleftmargin=1em,
  framexleftmargin=1em,
}

\usepackage{placeins}   

\begin{document}

\twocolumn[{%
    \renewcommand\twocolumn[1][]{#1}%
    \maketitle
    \begin{center}
        \captionsetup{type=figure}
        \includegraphics[width=\textwidth, trim={0 0 0 0}, clip]{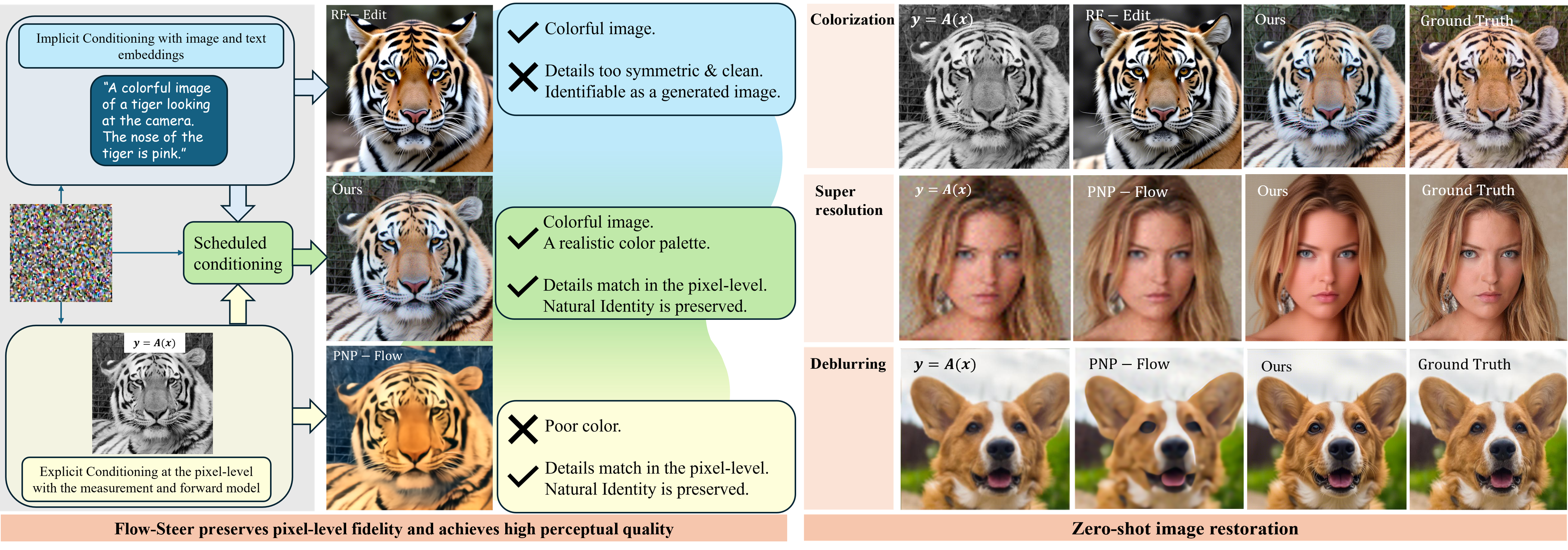}
        \caption{\small{Left: FlowSteer estimates a clean image $\mathbf{x}$ from a measurement $\mathbf{y}$ generated by a degradation operator $A$. It achieves both \textcolor{orange}{pixel-level fidelity by conditioning on the degradation model}, and \textcolor{blue}{high perceptual quality utilizing prompt and attention guidance}. Right: FlowSteer demonstrates advantage on colorization, super-resolution and deblurring tasks in both fidelity and perceptual quality.
        }}
        \label{fig:teaser}
    \end{center}

}]

\begin{abstract}
Flow-based text-to-image (T2I) models excel at prompt-driven image generation, but falter on Image Restoration (IR), often ``drifting away" from being faithful to the measurement. 
Prior work mitigate this drift with data-specific flows or task-specific adapters that are computationally heavy and not scalable across tasks. This raises the question ``Can't we efficiently manipulate the existing generative capabilities of a flow model?" 
To this end, we introduce \textbf{FlowSteer (FS)}, an operator-aware conditioning scheme that injects measurement priors along the sampling path,coupling a frozen flow's implicit guidance with explicit measurement constraints. 
Across super-resolution, deblurring, denoising, and colorization, FS improves measurement consistency and identity preservation in a strictly zero-shot setting—\textbf{no retrained models, no adapters}.
We show how the nature of flow models and their sensitivities to noise inform the design of such a scheduler.
FlowSteer, although simple, achieves a higher fidelity of reconstructed images, while leveraging the rich generative priors of flow models. 
All data and code will be publicly available \href{https://tharindu-nirmal.github.io/FlowSteer/}{in this link}.

\end{abstract}






\section{Introduction}
\label{sec:intro}

Over the past few years, diffusion models have demonstrated impressive generative capability~\cite{ho2020DDPM,song2020DDIM}, for many image editing~\cite{meng2022sdedit, avrahami2022blendeddiff,hertz2022_prompttoprompt,tumanyan2023plugdiff} and restoration tasks~\cite{saharia2022_SR3, chung2022DPS, kawar2022DDRM, song2023_ReSample, Blindfacerestoration}. Diffusion models, however, are known to have computational issues because one needs to run tens to hundreds of diffusion steps to obtain the result~\cite{dhariwal2021_diffusionbeatsgans,rombach2022high,shih2023parallel}. The more recently introduced flow models have shown initial success as an alternative to diffusion-based Text to Image/Video (T2I/T2V) models, where they can generate images with higher perceptual quality and aesthetics~\cite{yang2024_T2IRF_priors, rout2024_RFInversion, albergo2023_stochasticinterpolations, wan2025, Yuan_2025_NewtonGen}. However, flow models have not been widely adopted in image restoration tasks. If flow models are consistently faster and produce higher aesthetic quality, why are they not used in image restoration?


The biggest challenge of using a flow model for image restoration is the requirement for image fidelity. In image editing, the model can be more imaginative because the goal is not to preserve pixel-level fidelity~\cite{tumanyan2023_pnpdiffusion, wang2025point2pix, mou2024t2i, yang2023paint}. In image restoration, however, the forward image formation model provides a more restrictive constraint on what images are allowed to be generated. In theory, if a flow model is conditioned sufficiently, it is possible to obtain a restored image $\widehat{\mathbf{x}}$ that has sufficient fidelity, such that $\mathbf{\mathbf{y} \approx A\widehat{\mathbf{x}}}$, where $\mathbf{A}$ is the degradation formulation and $\mathbf{y}$ is the measurement of degraded image. If this can be achieved, then we will have both superior generative restoration and be much faster than the diffusion counterparts.

More recent image editing methods use implicit conditioning methods with embedded features~\cite{wang2024_RFEdit,wallace2023edict,meng2022sdedit}. First, a degraded input image is inverted into a noisy latent through flow inversion. At each step, features from underlying Diffusion Transformers (DiTs) are cached as layout guidance~\cite{kulikov2025flowedit,kim2025reflex_textguided, kim2025flowalign}. The inverted latent is then denoised by a prompt-guided Flux~\cite{FLUX2024,batifol2025_Flux} model, with selected steps fusing the cached features back into the trajectory. This approach only guides the overall layout of the input image, without any explicit conditioning on the pixel-level fidelity. Therefore, the output images have high perceptual quality and visual appeal, but the identity of the subject is lost. On the other hand, explicit pixel-level conditioning has been attempted through PNP-Flow~\cite{martin2024pnpflow} in every step of the denoising path. As a result, it is prone to excessive blur artifacts, and cannot achieve the rich colors, sharp textures, and high perceptual quality achievable with the generative priors.

In this paper, we present a flow-based image restoration method by making a critical observation: In Flow-based image restoration, the forward model cannot be \emph{uniformly} applied throughout the solution trajectory. 
Instead, we introduce a scheduler known as Flow-Steer(FS) to control the amount of forward model conditioning to be added to the solution trajectory. Specifically, no conditioning should be introduced in the beginning when the estimate is largely still in the latent space. Conditioning should be introduced toward the middle or latter stage of the solution trajectory when the estimate is becoming a meaningful image. Since Flow-Steer is a scheduler that can be added to any flow model, it offers a \emph{training-free} zero-shot upgrade of the flow-based image restoration at nearly no additional cost. To demonstrate this, we adapt a flow-based inversion-reconstruction pipeline. Shared attention features provide implicit conditioning combined with text prompts, whereas the Flow Steer schedule offers pixel-level conditioning.

In summary, our main contributions are threefold:
\begin{enumerate}
    \item We examine and explain the challenges of conditioning flow models towards pixel-level fidelity.
    \item We propose Flow-Steer, a scheduler that aims to mitigate the challenges by telling the flow model when to include forward model conditioning.
    \item We demonstrate that Flow-Steer is a universal scheme that can be adopted to a variety of image restoration tasks, including but not limited to deblurring, denoising, colorization, and super-resolution.
\end{enumerate}

\begin{figure*}[!t]
  \centering
  \includegraphics[width=\textwidth]{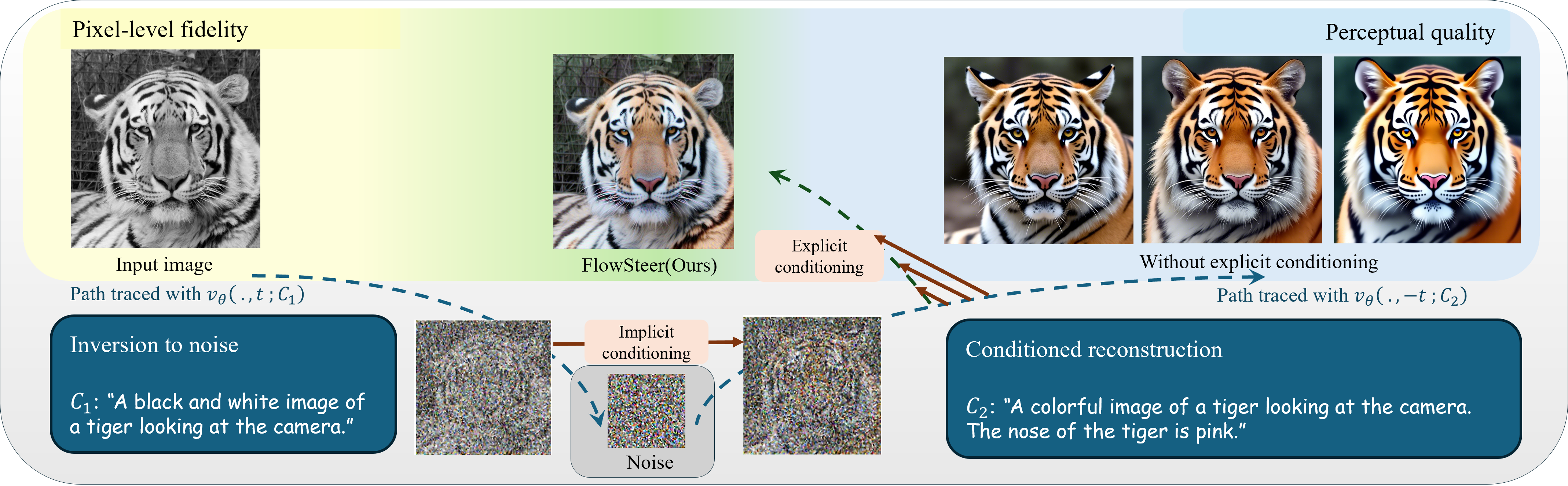}
  \caption{Image inversion(left) and reconstruction(right) paths of a flow model. A pre-trained flow trajectory $v_{\theta}$ is conditioned implicitly by language prompts ($C_1 , C_2)$ and feature sharing between paths. FlowSteer introduces an explicit conditioning schedule (center) to steer the reconstruction path toward high pixel-level fidelity without retraining the model.}
  \label{fig:flow_edit_landscape}
\end{figure*}


\begin{algorithm}[t]
\caption{Fidelity update in a diffusion path~\cite{wang2022DDNM}}
\label{alg:ddnm}
\begin{algorithmic}[1]
\State $\mathbf{x}_N \sim \N(\mathbf{0},\I)$
\For{$t = N, \ldots, 1$}
  \State $\mathbf{x}_{0\mid t} \gets \frac{1}{\sqrt{\bar{\alpha}_t}}
  \!\left(\mathbf{x}_t - \mathcal{Z}_{\theta}(\mathbf{x}_t,t)\sqrt{1-\bar{\alpha}_t}\right)$
  \Comment{denoise}
  
  \State $\widehat{\mathbf{x}}_{0\mid t} \gets \mathbf{A}^{\dagger}\mathbf{y}
  + \left(\I - \mathbf{A}^{\dagger}\mathbf{A}\right)\mathbf{x}_{0\mid t}$
  \Comment{fidelity update}
  
  \State $\mathbf{x}_{t-1} \sim \N\!\big(\mu_t(\mathbf{x}_t,\widehat{\mathbf{x}}_{0\mid t}),
  \sigma_t^2 \I\big)$ 
  \Comment{project back}
\EndFor
\State \Return $\mathbf{x}_0$
\end{algorithmic}
\end{algorithm}

\section{Related Works}
\label{sec:related_works}

\subsection{Image restoration with measurement fidelity}
\label{sec:related_restoration}

\textbf{Sampling diffusion paths.} Since the introduction of diffusion models~\cite{ho2020DDPM, song2020DDIM, song2020score, nichol2021improvedddpm} for image generation, people have quickly realized these models can be used in image restoration. Diffusion-based image restoration models~\cite{kawar2022DDRM,wang2022DDNM,choi2021ilvr_conditioningforddpm,kawar2021snips} pioneered operator-aware restorations. Every sample along the reconstruction path can be conditioned to have fidelity towards a noise-free forward model $\mathbf{y}=\mathbf{A}\mathbf{x}$ through a projection step. 
The core idea is to alternate between 1) a step in the direction of the log-likelihood $\nabla_\mathbf{x} \log \ p(\mathbf{y}|\mathbf{x})$ for higher fidelity, and 2) a denoising step or a step towards the log-prior $\nabla_\mathbf{x} \log \ p(x)$. Algorithm \ref{alg:ddnm} shows how the diffusion schedule of DDIM~\cite{song2020DDIM} was modified for image restoration through a $\nabla_\mathbf{x} \log \ p(\mathbf{y}|\mathbf{x})$ step by DDNM.~\cite{wang2022DDNM}.

\textbf{More updates along the sampling path.} 
Various additional improvements to the sampling path were introduced since then. DiffPIR~\cite{zhu2023DiffPIR} wraps a plug-and-play style update~\cite{venkatakrishnan2013pnp} inside each step along the diffusion sampling path. DPS~\cite{chung2022DPS} extends the work beyond the linear model, and relaxes a strict-measurement consistency update. This allows non-linear and noisy inverse problems to be solved with diffusion priors. Recently, DPIR~\cite{kong2025dualimgrestore} shows that using the both text prompts and the degraded image improves the reconstruction quality. RDMD~\cite{wang2025_RDMD} suggests a linearly weighted sum between a stochastic generative sample and a deterministic update can improve the perception-distortion trade-off. All of the above improvements can be considered as a ``scaffolding" to that of Algorithm\ref{alg:ddnm}.

\subsection{Flow-based image editing and restoration}
\label{sec:related_flowbased_edtitng}

\textbf{Flow models.}
Flow Matching (FM)~\cite{lipman2022_Flow} and Rectified Flow (RF)~\cite{liu2022_RectifiedFlow} reframed the generative modeling as learning velocity fields/ODE transports—aiming for linear sampling paths, fewer steps, and stronger quality~\cite{schusterbauer2025diff2flow}. Recent DiT/RF-Transformer systems~\cite{sauer2024fast} scaled these ideas for Text to Image(T2I), making flow-style models much faster during inference. (e.g., SD3/DiT~\cite{SD3,esser2024_DiT} and FLUX-style rectified-flow transformers~\cite{batifol2025_Flux}).

\textbf{Editing with rectified flows.}
RF models perform much better than their diffusion counterparts in image editing with T2I conditioning\cite{yang2024_T2IRF_priors,dalva2025fluxspace}. RF-Edit~\cite{wang2024_RFEdit} derives second-order samplers and share features between inversion and editing to reduce the drift. More work show fast, training-free inversion and editing in as few as ~8 steps, and edit specific regions of an image~\cite{deng2024_Fireflow,dalva2025fluxspace,jiao2025uniedit, rout2024_RFInversion}. However, these flow models are not common in image restoration for they are unable to produce consistent images with high fidelity to a measurement.

\textbf{Flow models for restoration.}
Works such as PnP-Flow~\cite{martin2024pnpflow}, D-Flow~\cite{ben2024_dflow}, and Flow Priors~\cite{zhang2024_flowpriors} demonstrate that flow models can be used in several image restoration tasks. However, these require a flow model to be separately trained for \textit{each} dataset. Thus, it is not transferrable to off-the shelf flow-based T2I models (such as flux~\cite{FLUX2024}), and hence is not scalable.

In this paper we show that pre-trained flow models can, in fact, be used for a range of image restoration tasks in a truly zero-shot manner, \textit{without} any retraining of the model. We find that flow models are sensitive to perturbations in conditioning, and that  the conditioning using the forward model should be introduced in middle or later stages of the reconstruction path.


\section{Flow Models and Restoration}
\label{sec:prelim}
We review Rectified Flow models and conditioned generation. We then see how inversion and reconstruction have been conditioned for image editing and their inherent limitations. Finally, we look at how the data fidelity term could be explicitly conditioned in an ideal flow model.

\subsection{Rectified Flow models}
\label{prelims_RF}

Flow models are trained to map a clean image $\mathbf{x}_0$ sampled from a clean distribution $\pi_0$, to a noise sample of the same dimensionality, $\mathbf{x}_1 \sim \pi_1$. 
For Rectified Flows (RF), intermediate samples along the path are linear interpolations between $\mathbf{x}_0$ and $\mathbf{x}_1$:
\begin{equation}
    \mathbf{x}_t = t\mathbf{x}_0 + (1-t)\mathbf{x}_1 \ , \ t \in [0,1] .
    \label{eq:rf-interpolation}
\end{equation}
 
The training scheme is to learn a velocity field $v_{\theta}(.)$, which is trained to predict the velocity between the two image distributions by minimizing

\begin{equation}
  \min_{\theta}\; \int_{0}^{1}
  \mathbb{E}\!\left[ \left\| (\mathbf{x}_1 - \mathbf{x}_0) - v_\theta(\mathbf{x}_t,t) \right\|_2^{2} \right] dt .
  \label{eq:rf-train}
\end{equation}

Given a sample of an intermediate image $\mathbf{x}_t$ and the time index $t \in (0,1)$, an ideally trained RF model should output the velocity that points to the desired clean image. For an incremental $d\mathbf{x}_t$, we can write $d\mathbf{x}_t = v_{\theta}(\mathbf{x}_t , t) dt$. Thereby, we have the Euler equation for solving the first-order ODE, implemented for a time schedule $t = {t_N, ... t_0}$:

\begin{equation}
  \mathbf{x}_{t_{i-1}} = \mathbf{x}_{t_i} + (t_{i-1}-t_i)\, v_\theta(\mathbf{x}_{t_i},t_i),
  \quad i\in\{N,\ldots,1\}.
  \label{eq:rf-euler}
\end{equation}


Assuming an ideal model $v_{\theta}$ that accurately estimates the velocity field conditioned on a text prompt $C$ and timestep $t$, Classifier-Free Guidance\cite{ho2022CFG} is further utilized to control the trajectory's drift $d\mathbf{x}_t$, to align with the target attributes specified by $C$.

\subsection{Image Inversion and Reconstruction}
\label{sec:Inversion_and_Reconstruction}
Given an input measurement $\mathbf{y}$ with degradation, the image editing framework first inverts it to a noise distribution $\pi_1$. In each step of inversion, the velocity field $v_{\theta}(.,t;\ C_1)$ is conditioned on a description $C_1$ of the degraded image. (See Figure \ref{fig:flow_edit_landscape}, left). Additionally, the attention maps at each step are cached to be fused with the reconstruction path.

Next, the inverted latent is denoised by sampling $v_{\theta}(.,-t; \ C_2)$, which is conditioned on a prompt $C_2$ that describes the target image. As illustrated on the right of Figure \ref{fig:flow_edit_landscape}, 
the discrete sampling steps formulate a trajectory from noise to clean image, where the cached attention maps are fused in the corresponding step. While this process steers the reconstruction toward a general manifold of high perceptual quality, the fidelity achieved by this standard inversion-and-reconstruction approach is insufficient for precise, high-quality image restoration.

\begin{figure}[!t]
  \centering
  \includegraphics[width=\textwidth]{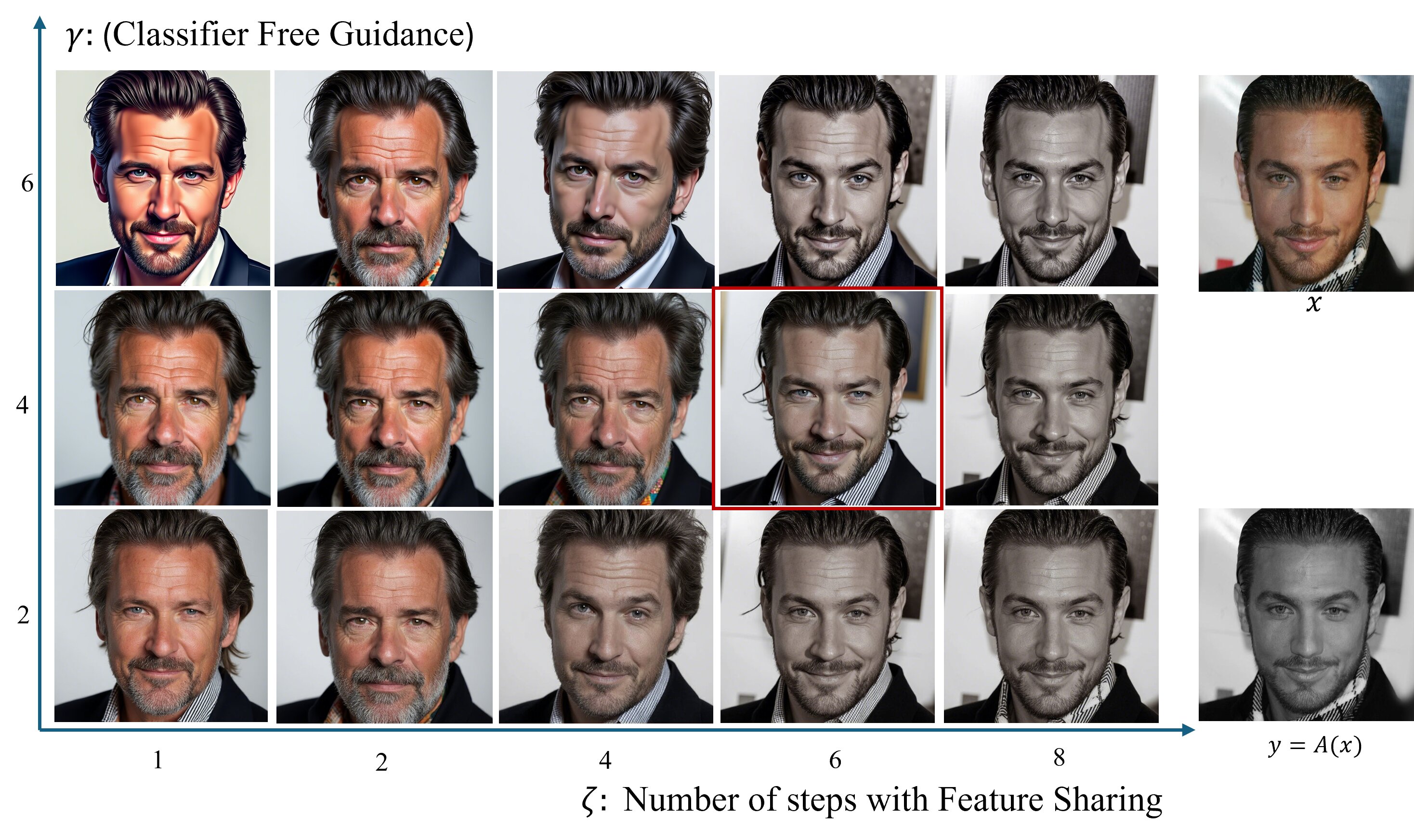}
  \caption{Increasing $\gamma$ enhances color saturation but reduces fidelity to the measurement $y$, while increasing $\zeta$ improves pixel-level consistency without recovering true color. An empirical pick from the grid search (red box) is still further from the ground truth $x$.}
  \label{fig:editing_limitation}
\end{figure}

\subsection{Limitations of Implicit Conditioning}
\label{prelims_limitations_implicit}
Fusing attention features from the inversion path into the reconstruction paths has its limitations.
Most pre-trained flow models such as Flux~\cite{FLUX2024} are trained in an encoded feature space mapped by VAE autoencoders~\cite{kingma2014vae,rombach2022high}. Thus, the entire flow algorithm runs on this embedding space, rather than in the pixel space. An attention map is a similarity of the tokens generated by $2\times2$ patches on these VAE embeddings, which is compressed from the pixel space (e.g, Flux uses $16\times$). Thus, fusing attention features only preserves coarse-level details without guidance on low-level texture within each token. Furthermore, these coarse features are not decoupled, and we cannot separate out the unwanted artifacts. Consider the example shown in figure~\ref{fig:editing_limitation}.  Having many steps ($\zeta$) that copy features from the inversion path tends to copy unwanted artifacts, leading the results to have a black-and-white color palette after reconstruction.

The other parameter used widely is the Classifier-Free-Guidance (CFG) \cite{ho2022CFG} parameter that controls the generative hallucinations of the flow model. This parameter $\gamma$ has to be tuned together with $\zeta$. Notice that even the best configuration of $(\gamma, \zeta)$, gives an image lacking in pixel-level fidelity. This motivates the explicit use of measurement $y$ and the degradation operator $\mathbf{A}$ for an explicit guidance towards the high fidelity region.

\begin{algorithm}[t]
\caption{Enforcing fidelity into an ideal flow path}
\label{alg:StF Ideal}
\begin{algorithmic}[1]
\State $\widehat{\mathbf{x}}_N \sim \pi_1 = \mathcal{N}(\mathbf{0}, \mathbf{I})$
\State $C \leftarrow \text{`` A colorful image of a tiger..." }$
\For{$i = N, \ldots, 1$}
  \State $\widehat{\mathbf{x}}_{t_i} = \widehat{\mathbf{x}}_{t_{i+1}} + v_{\theta}(\hat{\mathbf{x}}_{t_{i+1}}, t_i \ ; C) (t_{i+1} - t_i)$ 
  \State $\widehat{\mathbf{x}}_{0|t_i} = \frac{1}{(1-t_i) + \epsilon}\widehat{\mathbf{x}}_{t_i} -\frac{t_i}{(1-t_i) + \epsilon} \mathbf{\eta}$
  \Comment{denoise}
  \State $\bar{\mathbf{x}}_{0\mid t_i} \gets    \mathbf{A}^\dagger \mathbf{y} 
         + \big(\mathbf{I} - \mathbf{A}^\dagger \mathbf{A}\big)\,\mathbf{\widehat{x}}_{0\mid t_i} $ 
    \Comment{fidelity update}
  \State $\hat{\mathbf{x}}_{t_{i}} = (1-t_i) \mathbf{\bar{x}}_{0\mid t_i} + t\eta$ \Comment{project back}
\EndFor
\State \Return $\mathbf{x}_0$
\end{algorithmic}
\end{algorithm}


\subsection{Explicit Conditioning for Ideal Flows}
\label{prelims_explicit_conditioning}

Assuming a linear forward model $\mathbf{y}=\mathbf{Ax}$ is known, we can attempt for pixel level fidelity by adapting the ideas presented in section \ref{sec:related_restoration} by explicitly enforcing $\mathbf{A\bar{x}}_{0|t_i} = y$ as outlined in algorithm \ref{alg:StF Ideal}.

Suppose that at a time $t+1$ in the reconstruction path, we have a prediction for $\mathbf{x}_{t+1}$. This is line 4 in algorithm \ref{alg:StF Ideal}. To match with the framework of the null-space update in algorithm \ref{alg:ddnm}, we need to follow the forward model path defined for $v_{\theta}(.)$ to predict $\mathbf{\hat{x}_{0|t}}$.   
This is the denoising step (Step 3) of the algorithm \ref{alg:ddnm}, which can be adapted for flow using the forward model equation \ref{eq:rf-interpolation}. Assuming an ideally behaving flow model, the intermediate $\mathbf{x}_{t_i}$ lies in the linearly interpolated position $\widehat{\mathbf{x}}_{t_i} = t_i \mathbf{x}_1  + (1-t_i)\widehat{\mathbf{x}}_0$. 

Taking $\mathbf{x}_1$ to be Gaussian noise $\mathbf{\eta} \sim \mathcal{N}(0, \mathbf{I} )$, we get $\widehat{\mathbf{x}}_{0|t_i} = \frac{1}{(1-t_i) + \epsilon}\widehat{\mathbf{x}}_{t_i} -\frac{t_i}{(1-t_i) + \epsilon} \eta$. Notice how this can be interpreted as a denoising step $\widehat{\mathbf{x}}_{0|t} \gets D(\widehat{\mathbf{x}}_t)$. Next, the predicted clean image is updated in the null-space, followed by a re-projection back onto the intermediate step $t_i$.

Note that this analysis is valid if we have access to an ideal flow model $v_{\theta}$. Even with such a model, we require the $\epsilon$ term in the denominators during implementation for numerical stability.

\section{Steering Rectified Flows in the Wild}
\label{sec:practical_flow}

\subsection{Fragile, non-ideal flow models}
\textbf{Flow models are sensitive to noise.} Unlike diffusion models, flow models are not trained with a $\sigma_t$ schedule, and have no inherent ``buffer" for randomness in each step $t$. However, the following sources of noise $\eta$ are inevitable:
\begin{enumerate}
    \item No RF model will predict the true velocity in every time step. Thus $v_{\theta}(\widehat{\mathbf{x}}_{t_{i+1}}, t_i) = v_{ideal \ \theta}(\widehat{\mathbf{x}}_{t_{i+1}}, t_i) + \eta_1$ and \\ 
    $\mathbf{x}_{ t_i} = \mathbf{x}_{ t_i \ \text{ideal}} + \Delta t \eta_1$
    \item The forward and backward operators from $\mathbf{A}$ may not be fully accurate. For example, for colorization and superresolution, the pseudo-inverse operator $\mathbf{A^{\dagger}}$ is not exact. Hence $\widehat{\mathbf{x}}_{ t_i} \gets    \mathbf{A}^\dagger \mathbf{y} 
         +  \big(\mathbf{I} - \mathbf{A}^\dagger \mathbf{A}\big)\,\mathbf{\widehat{x}}_{t_i} + \eta_2$
    \item RF models are trained in an embedding space through autoencoders. These embeddings need to be converted back to a pixel space before the pixel-level fidelity is conditioned.
\end{enumerate}

The above sources of inherent randomness, added to numerical/precision errors, create a random noise at each step that we choose to enforce fidelity. Suppose that noise is characterized by a standard deviation $\eta_{\text{effective}}$. In diffusion models, the training scheme inherently has a random noise schedule $\sigma_t$. Therefore, as long as the hyper-parameters ensure $\eta_{\text{effective}} \ll \sigma_t $, such fidelity updates at every step do not pose problems.
In contrast, since the training scheme of flow models does not account for such randomness (see equation \ref{eq:rf-interpolation}), even a well-trained flow model with powerful priors cannot converge to desirable images with such noise-inducing fidelity updates on each step.
This is the reason for the fragility of using Diffusion-style updates on flow models, and we posit that this is a key challenge in the research community to adapt flow models for image restoration.

\subsection{Steering the Flow}
\label{sec:steering the flow}
As we shall see, there are some practical assumptions that will lead us to a schedule where an explicit update would not divert the flow unnecessarily. The only adjustment required is a parameter scheduler $\{ \lambda_i \}$, to achieve the explicit conditioning that we expected from Algorithm \ref{alg:StF Ideal}. Some practical decisions that lead to this scheduler are as follows:

\begin{figure}[!t]
  \centering
  \includegraphics[width=\textwidth]{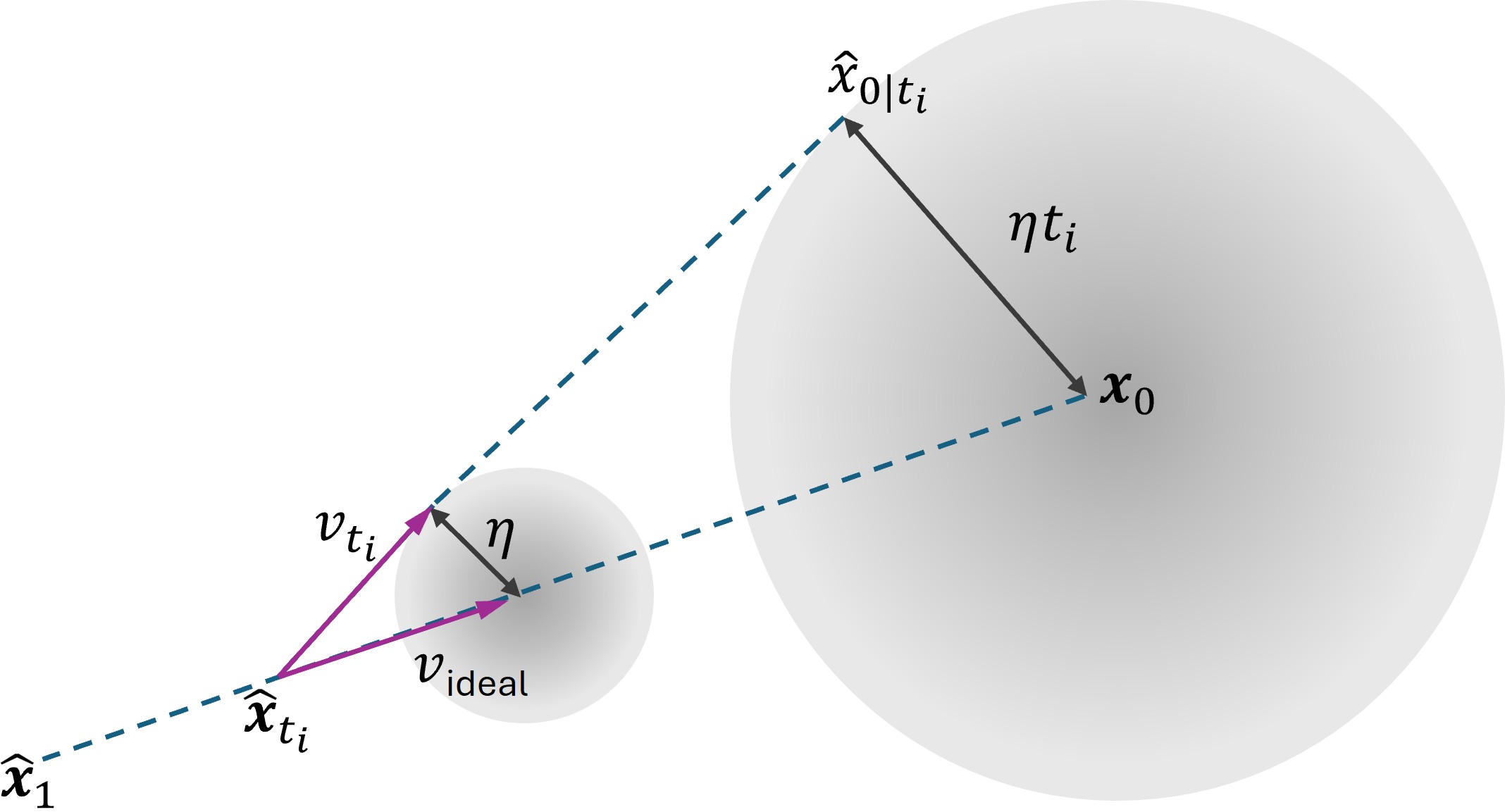}
  \caption{Linear projections of flow models exacerbate the errors from non-ideal velocities. We recommend to avoid projection operations to estimate $\mathbf{\widehat{x}}_{0|t_i}$ when $t_i \approx 1$.}
  \label{fig:avoidprojecting}
\end{figure}

\textbf{Avoid conditioning early on in the reconstruction.} Note that as a consequence of the non-ideal nature of the flow models, the noise gets scaled by a factor of $\Delta t$, and the estimation error for $\widehat{\mathbf{x}}_{0|t_i}$ scales by $t_i$ . A high-level view of this is shown in Figure \ref{fig:avoidprojecting}. 

\textbf{Avoid projecting to and from $\widehat{\textbf{x}}_{0|t}$ }.
When $t_i$ is close to 0, we can approximate the denoise step of algorithm \ref{alg:StF Ideal} to be $\widehat{\mathbf{x}}_{0|t_i} \approx \widehat{\mathbf{x}}_{t_i}$. This avoids linearly projecting to find an approximate $\mathbf{x}_{0|t_i}$. This results in the fidelity update enforcing  $\mathbf{A}\bar{\mathbf{x}}_{t} \equiv \mathbf{y}$ at the steps $t_i$ that we decide to condition the flow. This is a strict constraint to follow on each step, and makes intuitive sense: $\mathbf{y} = \mathbf{A}\mathbf{\widehat{x}}_t = \mathbf{A}\mathbf{\widehat{x}}_0 $ when $\mathbf{x}_t \simeq \mathbf{x}_0$. 

Therefore, we should avoid steps $5 \text{ and } 7$ in algorithm \ref{alg:StF Ideal}, and wait until $t_i$ is closer to $0$. One could also intuitively see from figure~\ref{fig:avoidprojecting}, that with such noise sensitivity, we should avoid more noise-inducing linear projections until necessary. Our ablations further validate these recommendations.

\begin{algorithm}[t]
\caption{FlowSteer: for a Non-ideal Flow model}
\label{alg:StF practical}
\begin{algorithmic}[1]
\State $\mathbf{\widehat{z}}_N \sim \pi_1 = \mathcal{N}(\mathbf{0}, \mathbf{I})$
\State $C \leftarrow \text{`` A colorful image of a tiger..." }$
\For{$i = N, \ldots, 1$}
  \State $\hat{\mathbf{z}}_{t_i} = \hat{\mathbf{z}}_{t_{i+1}} + v_{\theta}(\hat{\mathbf{z}}_{t_{i+1}}, t_i \ ;C) (t_{i+1} - t_i)$
  \State $\widehat{\mathbf{x}}_{t_i} = \text{Decoder}( \widehat{\mathbf{z}}_{t_i} ) $
  \State $ \quad \text{if} \quad \lambda_{i} > 0$:
  \State $ \quad \widehat{\mathbf{x}}_{ t_i} \gets    \mathbf{A}^\dagger \mathbf{y} 
         + \lambda_{i} \big(\mathbf{I} - \mathbf{A}^\dagger \mathbf{A}\big)\,\mathbf{\widehat{x}}_{t_i} $ \Comment{fidelity update}
  \State $\widehat{\mathbf{z}}_{t_i} = \text{Encoder}( \widehat{\mathbf{x}}_{t_i} ) $
\EndFor
\State \Return $\text{Decoder}(\mathbf{z}_0)$
\end{algorithmic}
\end{algorithm}

\subsection{A sparse update scheduler}
The above two suggestions lead us to design a scheduler that decides which steps during the reconstruction path, which enforces $\mathbf{A}\widehat{\mathbf{x}}_{t_i} \equiv \mathbf{y}$, achieve desirable fidelity. To control this, we propose using the fidelity update step with a sparse parameter schedule $\{\lambda_{i}\}$, for which the fidelity update triggers only if $\lambda_i >0$. Furthermore, since available RF models operate in an embedding space, a Decoder-Encoder wrapper is needed to convert these embeddings into pixel space. The complete procedure is summarized in Algorithm \ref{alg:StF practical}.

\subsection{Our baseline flow model and schedule recommendations}
\textbf{The restoration pipeline.} We first convert our input image into a latent space using the ViT autoencoders. These latents are inverted into a noisy latent distribution through flow inversion, and we cache the attention map during inversion as layout guidance. Then, the inverted latent is denoised with multi-step inference of the Flux model. Our FlowSteer schedule decides the steps in which we merge both the layout guidance from the attention map and pixel-wise guidance from the degraded prior. In these steps, the latents are decoded back to image space, and the fidelity update is performed.

\textbf{The flow model}. We select the Flux-dev model~\cite{FLUX2024} as our pre-trained RF model, with inversion and reconstruction paths having $N=30$ steps each. We then design our image inversion and reconstruction pipeline as described in section \ref{sec:Inversion_and_Reconstruction}. For each input image, an input caption $C_1$ describes the image with its degradation, and the target caption $C_2$ describes the ideally restored image. We cache the attention maps of each inversion step for copying in the reconstruction path, specified by $\zeta$. The hyperparameters $\gamma \text{ and } \zeta$ (as described in Section~\ref{prelims_limitations_implicit}) are chosen so that the best-available trade-off between fidelity and perceptual quality is obtained, and is kept at $(4,4)$ for all experiments. This forms our implicit conditioning scheme, on top of which FlowSteer is applied.

\textbf{FlowSteer scheduler.}
To improve fidelity and explicitly condition the forward model, we design the FS scheduler $\{ \lambda_i \}$ so that a set of validation images achieves a desirable trade-off between fidelity and reconstruction quality. At its simplest form, the schedule will be a rectangular window, with three parameters $i_{\text{start}}, i_{\text{stop}}, h$ defining the schedule: 
\begin{equation}
    \lambda_i = 
    \begin{cases}
        h & \text{if} \quad  i_{\text{start}} \le i \le i_{\text{stop}} \ ,\\
        0  & \text{otherwise}.
    \end{cases}
\end{equation}

\textbf{Schedule recommendations.} In general, we recommend that the schedule start enforcing fidelity between 50\%-90\% of the total steps. The intuition for this is based on how image generation proceeds from coarse to fine details along the reconstruction path, as discussed in recent work~\cite{hertz2022_prompttoprompt, park2023_understandinglatentsindiffusion,saichandran2025_progressivepromptdealing}. In the early steps, the color palette is formed, and the foreground and background are separated. Towards the end, finer details and textures are formed. Conditioning too late towards the end risks unwanted hallucinations. Conditioning too early introduces a higher effective noise (as described in Section \ref{sec:steering the flow}) and directs the restoration toward a noisy output. Depending on the reconstruction task, $(i_{\text{start}}, i_{\text{stop}})$ can be further fine-tuned for improved performance. The relative strength of the fidelity update can also be controlled by $h$ to mitigate unwanted artifacts from the noisy conditioning step. Refer to ablations in Section \ref{ablation} for the effects of $(i_{\text{start}}, i_{\text{stop}})$ and other alternatives.



\begin{figure}[!t]
  \centering
  \includegraphics[width=\textwidth]{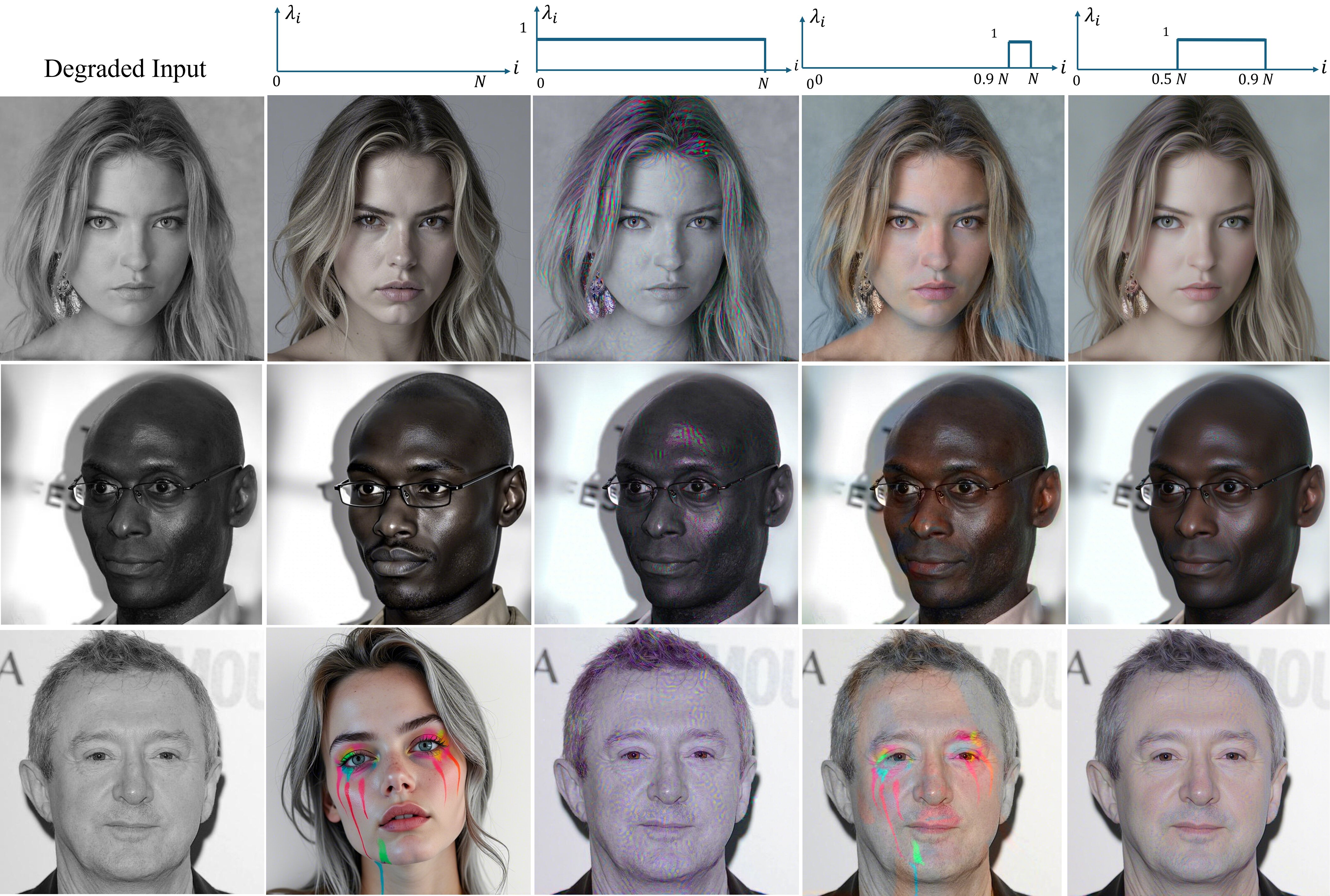}
  \caption{Colorizing with different $\{ \lambda_i\}$ schedules on our baseline RF model. No explicit conditioning creates hallucinations and loses identity. Having a constant conditioning with $\lambda = 1$ introduces undesired noise.}
  \label{fig:lambda_schedule}
\end{figure}

\section{Experiments}
\begin{figure*}
\centering
\begingroup
\setlength{\tabcolsep}{0pt} 
\renewcommand{\arraystretch}{0.1} 
\begin{tabular}{cccccccc} 
    
    \rotatebox[origin=l]{90}{\textbf{Colorization}} &

    \begin{tikzpicture}
        \node[anchor=south west, inner sep=0] (image) at (0,0) {\includegraphics[width=0.140\textwidth]{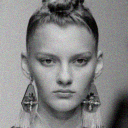}};
    \end{tikzpicture} &
    \begin{tikzpicture}
        \node[anchor=south west, inner sep=0] (image) at (0,0) {\includegraphics[width=0.140\textwidth]{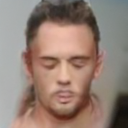}};
        \begin{scope}[x={(image.south east)}, y={(image.north west)}] \draw[red, thick, -{Latex[length=2mm]}] (0.9,0.15) -- (0.8,0.2); \end{scope}
    \end{tikzpicture} &
    \begin{tikzpicture}
        \node[anchor=south west, inner sep=0] (image) at (0,0) {\includegraphics[width=0.140\textwidth]{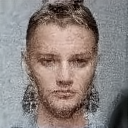}};
        \begin{scope}[x={(image.south east)}, y={(image.north west)}] \draw[red, thick, -{Latex[length=2mm]}] (0.9,0.15) -- (0.8,0.2); \end{scope}
    \end{tikzpicture} &
    \begin{tikzpicture}
        \node[anchor=south west, inner sep=0] (image) at (0,0) {\includegraphics[width=0.140\textwidth]{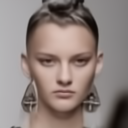}};
        \begin{scope}[x={(image.south east)}, y={(image.north west)}] \draw[red, thick, -{Latex[length=2mm]}] (0.9,0.15) -- (0.8,0.2); \end{scope}
    \end{tikzpicture} &
    \begin{tikzpicture}
        \node[anchor=south west, inner sep=0] (image) at (0,0) {\includegraphics[width=0.140\textwidth]{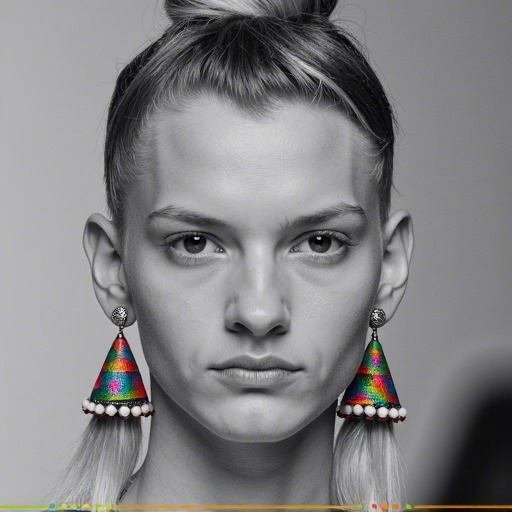}};
        \begin{scope}[x={(image.south east)}, y={(image.north west)}] \draw[red, thick, -{Latex[length=2mm]}] (0.9,0.15) -- (0.8,0.2); \end{scope}
    \end{tikzpicture} &
    \begin{tikzpicture}
        \node[anchor=south west, inner sep=0] (image) at (0,0) {\includegraphics[width=0.140\textwidth]{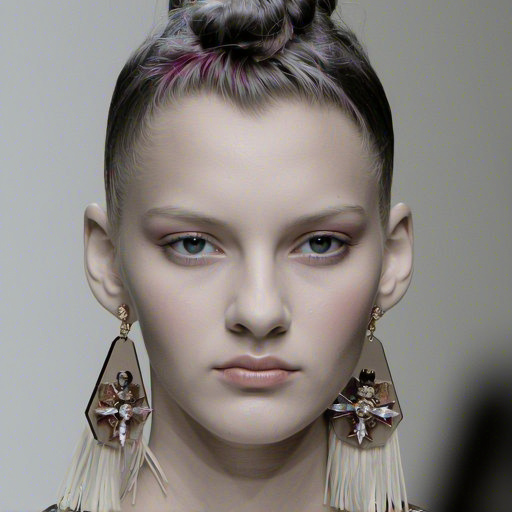}};
        \begin{scope}[x={(image.south east)}, y={(image.north west)}] \draw[green, thick, -{Latex[length=2mm]}] (0.9,0.15) -- (0.8,0.2); \end{scope}
    \end{tikzpicture} &
    \begin{tikzpicture}
        \node[anchor=south west, inner sep=0] (image) at (0,0) {\includegraphics[width=0.140\textwidth]{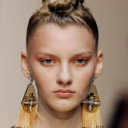}};
    \end{tikzpicture} \\

    \rotatebox[origin=l]{90}{\textbf{Super-res}} &
    \begin{tikzpicture}
        \node[anchor=south west, inner sep=0] (image) at (0,0) {\includegraphics[width=0.140\textwidth]{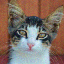}};
    \end{tikzpicture} &
    \begin{tikzpicture}
        \node[anchor=south west, inner sep=0] (image) at (0,0) {\includegraphics[width=0.140\textwidth]{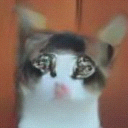}};
        \begin{scope}[x={(image.south east)}, y={(image.north west)}] \draw[red, thick, -{Latex[length=2mm]}] (0.7,0.15) -- (0.6,0.2); \end{scope}
    \end{tikzpicture} &
    \begin{tikzpicture}
        \node[anchor=south west, inner sep=0] (image) at (0,0) {\includegraphics[width=0.140\textwidth]{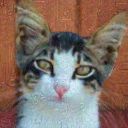}};
        \begin{scope}[x={(image.south east)}, y={(image.north west)}] \draw[red, thick, -{Latex[length=2mm]}] (0.7,0.15) -- (0.6,0.2); \end{scope}
    \end{tikzpicture} &
    \begin{tikzpicture}
        \node[anchor=south west, inner sep=0] (image) at (0,0) {\includegraphics[width=0.140\textwidth]{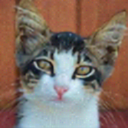}};
        \begin{scope}[x={(image.south east)}, y={(image.north west)}] \draw[red, thick, -{Latex[length=2mm]}] (0.7,0.15) -- (0.6,0.2); \end{scope}
    \end{tikzpicture} &
    \begin{tikzpicture}
        \node[anchor=south west, inner sep=0] (image) at (0,0) {\includegraphics[width=0.140\textwidth]{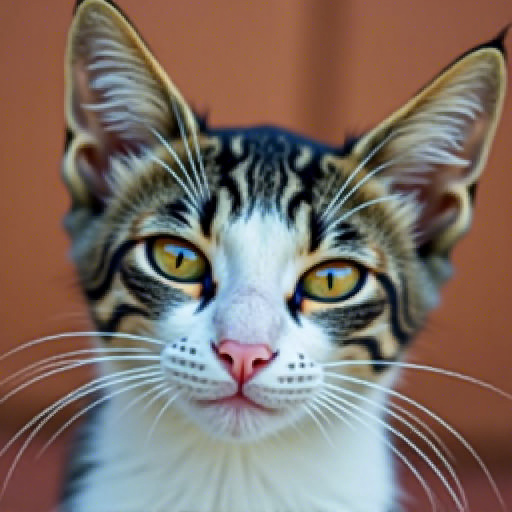}};
        \begin{scope}[x={(image.south east)}, y={(image.north west)}] \draw[red, thick, -{Latex[length=2mm]}] (0.7,0.15) -- (0.6,0.2); \end{scope}
    \end{tikzpicture} &
    \begin{tikzpicture}
        \node[anchor=south west, inner sep=0] (image) at (0,0) {\includegraphics[width=0.140\textwidth]{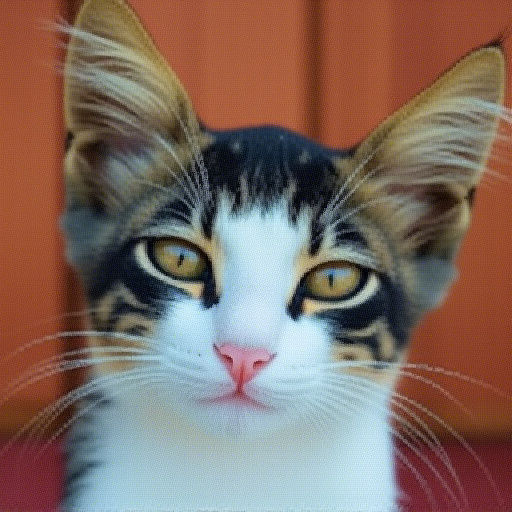}};
        \begin{scope}[x={(image.south east)}, y={(image.north west)}] \draw[green, thick, -{Latex[length=2mm]}] (0.7,0.15) -- (0.6,0.2); \end{scope}
    \end{tikzpicture} &
    \begin{tikzpicture}
        \node[anchor=south west, inner sep=0] (image) at (0,0) {\includegraphics[width=0.140\textwidth]{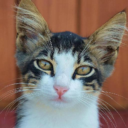}};
    \end{tikzpicture} \\

    
    \rotatebox[origin=l]{90}{\textbf{Denoising}} &
    \begin{tikzpicture}
        \node[anchor=south west, inner sep=0] (image) at (0,0) {\includegraphics[width=0.140\textwidth]{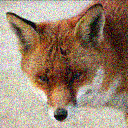}};
    \end{tikzpicture} &
    \begin{tikzpicture}
        \node[anchor=south west, inner sep=0] (image) at (0,0) {\includegraphics[width=0.140\textwidth]{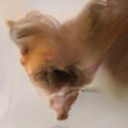}};
        \begin{scope}[x={(image.south east)}, y={(image.north west)}] \draw[red, thick, -{Latex[length=2mm]}] (0.7,0.3) -- (0.6,0.35); \end{scope}
    \end{tikzpicture} &
    \begin{tikzpicture}
        \node[anchor=south west, inner sep=0] (image) at (0,0) {\includegraphics[width=0.140\textwidth]{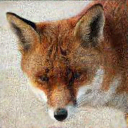}};
        \begin{scope}[x={(image.south east)}, y={(image.north west)}] \draw[red, thick, -{Latex[length=2mm]}] (0.7,0.3) -- (0.6,0.35); \end{scope}
    \end{tikzpicture} &
    \begin{tikzpicture}
        \node[anchor=south west, inner sep=0] (image) at (0,0) {\includegraphics[width=0.140\textwidth]{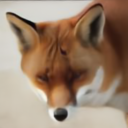}};
        \begin{scope}[x={(image.south east)}, y={(image.north west)}] \draw[red, thick, -{Latex[length=2mm]}] (0.7,0.3) -- (0.6,0.35); \end{scope}
    \end{tikzpicture} &
    \begin{tikzpicture}
        \node[anchor=south west, inner sep=0] (image) at (0,0) {\includegraphics[width=0.140\textwidth]{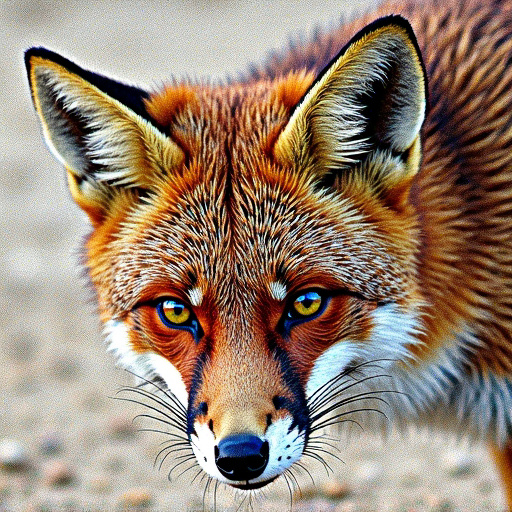}};
        \begin{scope}[x={(image.south east)}, y={(image.north west)}] \draw[red, thick, -{Latex[length=2mm]}] (0.7,0.3) -- (0.6,0.35); \end{scope}
    \end{tikzpicture} &
    \begin{tikzpicture}
        \node[anchor=south west, inner sep=0] (image) at (0,0) {\includegraphics[width=0.140\textwidth]{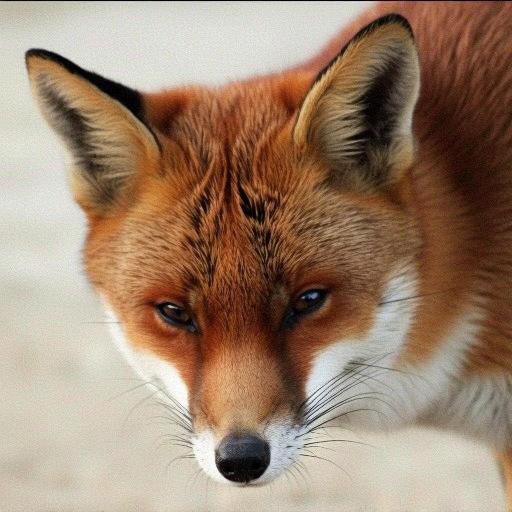}};
        \begin{scope}[x={(image.south east)}, y={(image.north west)}] \draw[green, thick, -{Latex[length=2mm]}] (0.7,0.3) -- (0.6,0.35); \end{scope}
    \end{tikzpicture} &
    \begin{tikzpicture}
        \node[anchor=south west, inner sep=0] (image) at (0,0) {\includegraphics[width=0.140\textwidth]{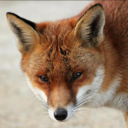}};
    \end{tikzpicture} \\


    \rotatebox[origin=l]{90}{\textbf{Deblurring}} &

    \begin{tikzpicture}
        \node[anchor=south west, inner sep=0] (image) at (0,0) {\includegraphics[width=0.140\textwidth]{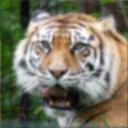}};
    \end{tikzpicture} &
    \begin{tikzpicture}
        \node[anchor=south west, inner sep=0] (image) at (0,0) {\includegraphics[width=0.140\textwidth]{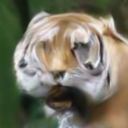}};
        \begin{scope}[x={(image.south east)}, y={(image.north west)}] \draw[red, thick, -{Latex[length=2mm]}] (0.7,0.15) -- (0.6,0.2); \end{scope}
    \end{tikzpicture} &
    \begin{tikzpicture}
        \node[anchor=south west, inner sep=0] (image) at (0,0) {\includegraphics[width=0.140\textwidth]{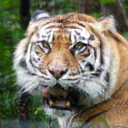}};
        \begin{scope}[x={(image.south east)}, y={(image.north west)}] \draw[red, thick, -{Latex[length=2mm]}] (0.7,0.15) -- (0.6,0.2); \end{scope}
    \end{tikzpicture} &
    \begin{tikzpicture}
        \node[anchor=south west, inner sep=0] (image) at (0,0) {\includegraphics[width=0.140\textwidth]{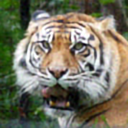}};
        \begin{scope}[x={(image.south east)}, y={(image.north west)}] \draw[red, thick, -{Latex[length=2mm]}] (0.7,0.15) -- (0.6,0.2); \end{scope}
    \end{tikzpicture} &
    \begin{tikzpicture}
        \node[anchor=south west, inner sep=0] (image) at (0,0) {\includegraphics[width=0.140\textwidth]{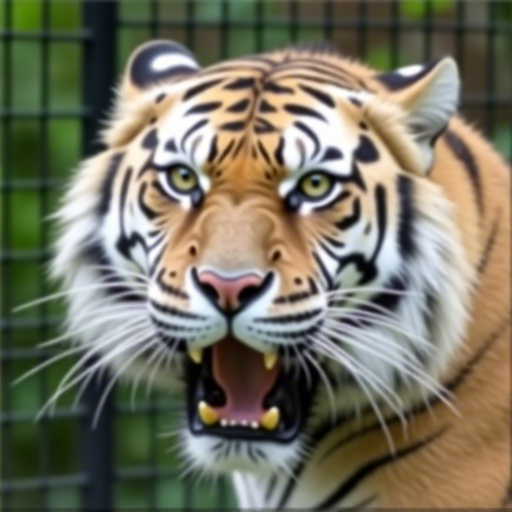}};
        \begin{scope}[x={(image.south east)}, y={(image.north west)}] \draw[red, thick, -{Latex[length=2mm]}] (0.7,0.15) -- (0.6,0.2); \end{scope}
    \end{tikzpicture} &
    \begin{tikzpicture}
        \node[anchor=south west, inner sep=0] (image) at (0,0) {\includegraphics[width=0.140\textwidth]{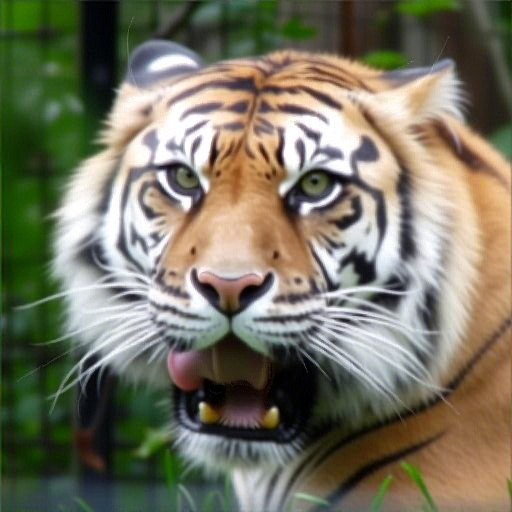}};
        \begin{scope}[x={(image.south east)}, y={(image.north west)}] \draw[green, thick, -{Latex[length=2mm]}] (0.7,0.15) -- (0.6,0.2); \end{scope}
    \end{tikzpicture} &
    \begin{tikzpicture}
        \node[anchor=south west, inner sep=0] (image) at (0,0) {\includegraphics[width=0.140\textwidth]{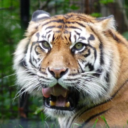}};
    \end{tikzpicture} \\

    \rule{0pt}{4ex}  & Degraded  & D-Flow~\cite{ben2024_dflow} & {FlowPriors~\cite{zhang2024_flowpriors}} & PnP-Flow~\cite{martin2024pnpflow} & RFEdit~\cite{wang2024_RFEdit} & Ours & Clean \\
\end{tabular}
\endgroup
\caption{Qualitative comparison of flow-based methods. Columns 2-4 are image restoration models, and column 5 is an image editing model. The rows show four degradations: tasks focused on information reconstruction (colorization and $4\times$ super-resolution) and tasks focused on corruption removal (denoising and deblurring). FlowSteer removes degradations, and details are generated without losing the identity of the subject.}
\label{fig:qualitative_comparison}
\end{figure*}

In this section we evaluate FlowSteer on image restoration tasks. Section \ref{implementation details} presents implementation details, Section \ref{sec:comparisons} compares our framework with other comparable methods, and Section \ref{ablation} discusses ablation studies.

\subsection{Implementation details}
\label{implementation details}  

\textbf{Forward models.} We select the restoration restoration tasks: colorization, denoising, deblurring, and 4x super resolution. Each forward model is modeled as an operator $\textbf{A}$, and assigned a pesudo-inverse operator $\mathbf{A}^\dagger$ to implement algorithm \ref{alg:StF practical}. Linear transforms are assumed following Wang \etal~\cite{wang2022DDNM} for colorization and super resolution.  For Deblurring, we follow Martin \etal~\cite{martin2024pnpflow} with a $61 \times 61 $-pixel gaussian kernel with blur $\sigma_b = 3.0$ for $\mathbf{A}$. A Wiener deconvolution operator~\cite{gonzalez2018digital,hansen2006deblurring} with $\Lambda_{\text{Wiener}}=0.1$ is used for $\mathbf{A}^\dagger$. For denoising, the forward model is an additive Gaussian noise of $\sigma_g =0.2$, and we set $\mathbf{A} = \mathbf{A}^\dagger = \mathbf{I}$ for the fidelity update. Refer to the supplementary for more details.

\textbf{Data and metrics.} 
A set of 100 sampled images from AFHQ~\cite{choi2020stargan_AFHQ}, CelebAHQ~\cite{liu2015deep_CelebA,karras2017progressive_CelebAHQ} data is selected as our test data. A sample of 20 images from the same source is the validation split to tune hyper-parameters and run ablations. This forms a collection of human faces, pets, and wild animals. Our evaluation focuses on pixel-level fidelity and perceptual quality. PSNR, SSIM metrics measure the fidelity/ the consistency of the generated image to the forward model. To report the perceptual similarity between the predicted and ground truth images, we use LPIPS~\cite{Zhang_2018_PerceptualLPIPS} and the Cosine similarity between CLIP~\cite{radford2021learning_CLIP} embeddings. For Colorization, we first implement a histogram matching algorithm for each color channel to ensure that global effects such as brightness, saturation shifts, and contrast doesn't effect the metrics.

\subsection{Comparisons}
\label{sec:comparisons}
We compare the restoration capability of FlowSteer with other flow models: OT-ODE~\cite{pokle2024_OTODE}, D-Flow~\cite{ben2024_dflow}, Flow-Priors~\cite{zhang2024_flowpriors}, FlowChef~\cite{patel2024flowchef}, and PNP-Flow~\cite{martin2024pnpflow}. These methods induce a fidelity update in each step and, therefore, result in higher artifacts in the final restoration.
To reproduce PNP-Flow~\cite{martin2024pnpflow}, we use the separate flow models for humans and animals as provided by the authors. However, for our method, we use \textit{the same flux model for all types of data}, and show that a large pretrained T2I model, such as Flux, can be steered for faithful image restoration. 

Since models with an inversion-reconstruction pipeline have not been used for image restoration, we reproduce RF-Edit~\cite{wang2024_RFEdit}, tune the corresponding $(\gamma, \zeta)$ parameters through a grid-search, and keep them constant throughout the test data. This baseline demonstrates that implicit conditioning loses the identity of the image, and that FlowSteer can achieve high pixel-level fidelity while maintaining a high visual quality.
Refer supplementary for more details.

\textbf{Fidelity and Perceptual quality.} FlowSteer preserves the perceptual qualities of the pre-trained T2I flux model (such as rich colors, fine details, and textures), while also achieving a high pixel-level fidelity with the degraded measurement. It enhances degraded details, without losing the identity of the subject. In the qualitative results of Figure \ref{fig:qualitative_comparison}, we see that characteristic facial features, whiskers, teeth,fur ...etc are enhanced, without changing the identity of the subject.
This is reflected quantitatively in Table \ref{tab:baselines}. 

\begin{table*}[ht]
\centering
\small
\resizebox{\textwidth}{!}{
\begin{tabular}{
  l
  S[table-format=2.4] S[table-format=1.4] S[table-format=1.4] S[table-format=1.4]
  S[table-format=2.4] S[table-format=1.4] S[table-format=1.4] S[table-format=1.4]
  S[table-format=2.4] S[table-format=1.4] S[table-format=1.4] S[table-format=1.4]
  S[table-format=2.4] S[table-format=1.4] S[table-format=1.4] S[table-format=1.4]
}
\toprule
\multirow{3}{*}{Methods}
& \multicolumn{4}{c}{Colorization}
& \multicolumn{4}{c}{Super resolution}
& \multicolumn{4}{c}{Deblurring}
& \multicolumn{4}{c}{Denoising}
\\
\cmidrule(lr){2-5}\cmidrule(lr){6-9}\cmidrule(lr){10-13}\cmidrule(lr){14-17}
& {PSNR$\uparrow$} & {SSIM$\uparrow$} & {LPIPS$\downarrow$} & {CLIP$\uparrow$}
& {PSNR$\uparrow$} & {SSIM$\uparrow$} & {LPIPS$\downarrow$} & {CLIP$\uparrow$}
& {PSNR$\uparrow$} & {SSIM$\uparrow$} & {LPIPS$\downarrow$} & {CLIP$\uparrow$}
& {PSNR$\uparrow$} & {SSIM$\uparrow$} & {LPIPS$\downarrow$} & {CLIP$\uparrow$}
\\
\midrule
\textit{Reference}
& 27.1891 & 0.8333 & 0.2334 & 0.5807
& 26.7337 & 0.7829 & 0.2381 & 0.4380
& 28.2371 & 0.7586 & 0.2821 & 0.3095
& 22.7706 & 0.4399 & 0.4427 & 0.3178
\\
\midrule
D-Flow~\cite{ben2024_dflow}
& 18.5295 & 0.5554 & 0.4977 & 0.2724
& 23.4039 & 0.6236 & 0.4132 & 0.2984
& 22.2396 & 0.6316 & 0.4104 & 0.2798
& 19.0651 & 0.5232 & 0.4840 & 0.2451
\\
OT-ODE~\cite{pokle2024_OTODE}
& \multicolumn{4}{c}{N/A}
& 29.1072 & 0.8296 & 0.1994 & 0.6033
& 31.0467 & 0.8612 & 0.1984 & 0.5908
& 28.8276 & 0.8123 & 0.2505 & 0.4416
\\
Flow-Priors~\cite{zhang2024_flowpriors}
& 21.7889 & 0.5716 & 0.4639 & 0.4808
& 27.6858 & 0.7151 & 0.3036 & 0.4724
& 30.4326 & 0.8350 & 0.2264 & 0.5535
& 28.8047 & 0.7645 & 0.2770 & 0.4479
\\
FlowChef~\cite{patel2024flowchef}
& 18.5834 & 0.5537 & 0.4640 & 0.1857 
& \BB{31.3876} & 0.8430 & 0.1849 & 0.5290 
& \BB{32.8454} & \BB{0.8929} & 0.2093 & 0.3065 
& 29.6792 & \BB{0.8803} & 0.3421 & 0.2711 
\\
PnP-flow~\cite{martin2024pnpflow}
& \BB{27.1620} & \BB{0.8640} & \BB{0.2830} & 0.3482
& 31.2073 & \BB{0.8753} & \BB{0.1755} & 0.3938
& 32.7392 & 0.8840 & \BB{0.1728} & 0.4597
& \BB{30.5899} & 0.8733 & \BB{0.2207} & 0.5103
\\
RFEdit~\cite{wang2024_RFEdit}
& 20.3255 & 0.6602 & 0.3648 & \RB{0.8703}
& 22.3243 & 0.7362 & 0.2822 & \RB{0.7863}
& 22.9813 & 0.7431 & 0.2750 & \BB{0.7914}
& 17.0219 & 0.4438 & 0.4545 & \RB{0.8570}
\\
Ours
& \RB{27.4214} & \RB{0.8696} & \RB{0.2081} & \BB{0.7734}
& \RB{32.8552} & \RB{0.9022} & \RB{0.1700} & \BB{0.6714}
& \RB{32.8749} & \RB{0.9052} & \RB{0.1486} & \RB{0.8177}
& \RB{32.2125} & \RB{0.8924} & \RB{0.1822} & \BB{0.7679}
\\
\bottomrule
\end{tabular}}
\caption{Quantitative comparison with flow-based restoration methods. \textit{Reference} uses the degraded image $y$. FlowSteer has high perceptual quality, and high pixel-level fidelity. We highlight the \colorbox{red!20}{best} and \colorbox{blue!20}{second-best} per metric.}
\label{tab:baselines}
\end{table*}




\begin{table*}[t]
\centering
\scriptsize
\setlength{\tabcolsep}{3.5pt}
\renewcommand{\arraystretch}{0.92}
\sisetup{table-number-alignment=center,round-mode=places,round-precision=4}
\begin{tabular}{
  @{}l
  l S[table-format=2.4] S[table-format=1.4]
  l S[table-format=2.4] S[table-format=1.4]@{}
}
\toprule
\multirow{2}{*}{Degradation}
& \multicolumn{3}{c}{General schedule}
& \multicolumn{3}{c}{Task-specific fine-tuned schedule} \\
\cmidrule(lr){2-4}\cmidrule(lr){5-7}
& Params $(i_{\text{start}},i_{\text{end}},h)$
& {PSNR$\uparrow$} & {LPIPS$\downarrow$}
& Params $(i_{\text{start}},i_{\text{step}},i_{\text{end}},h_1,h_2)$
& {PSNR$\uparrow$} & {LPIPS$\downarrow$} \\
\midrule
Colorization
& $(0.5N,0.9N,1)$ & 22.5841 & 0.3065
& $(0.4N,0.50N,0.95N,1,0.3)$ & \bfseries 27.4214 & \bfseries 0.2081 \\
Super-resolution
& $(0.5N,0.9N,1)$ & 22.8262 & 0.3753
& $(0.5N,0.70N,0.85N,1,0.5)$ & \bfseries 32.8552 & \bfseries 0.1700 \\
Deblurring
& $(0.5N,0.9N,1)$ & 23.1362 & 0.3959
& $(0.7N,0.80N,0.90N,1,0.3)$ & \bfseries 32.8749 & \bfseries 0.1486 \\
Denoising
& $(0.5N,0.9N,1)$ & 22.4335 & 0.4677
& $(0.5N,0.75N,0.95N,1,0.5)$ & \bfseries 30.3822 & \bfseries 0.2313 \\
\bottomrule
\end{tabular}
\caption{Comparison of general and fine-tuned FlowSteer schedule settings across degradation tasks.}
\label{tab:finetuned_schedule}
\vspace{-2mm}
\end{table*}
\begin{table}[t]
\centering
\footnotesize
\setlength{\tabcolsep}{3.5pt}
\renewcommand{\arraystretch}{0.95}
\sisetup{table-number-alignment=center,round-mode=places,round-precision=4}
\begin{tabular}{
  l
  S[table-format=2.4] S[table-format=1.4]
  S[table-format=2.4] S[table-format=1.4]
}
\toprule
\multirow{2}{*}{Degradation}
& \multicolumn{2}{c}{with $\widehat{\mathbf{x}}_{0|t}\!\leftarrow\!\widehat{\mathbf{x}}_t$}
& \multicolumn{2}{c}{w/o $\widehat{\mathbf{x}}_{0|t}\!\leftarrow\!\widehat{\mathbf{x}}_t$} \\
& {PSNR$\uparrow$} & {LPIPS$\downarrow$}
& {PSNR$\uparrow$} & {LPIPS$\downarrow$} \\
\midrule
Colorization        & 11.8157 & 0.6256 & \bfseries 27.4214 & \bfseries 0.2081 \\
Super-resolution    & 12.0936 & 0.6168 & \bfseries 32.8552 & \bfseries 0.1700 \\
Deblurring          & 12.1171 & 0.6228 & \bfseries 32.8749 & \bfseries 0.1486 \\
Denoising           & 12.1006 & 0.6467 & \bfseries 30.3822 & \bfseries 0.2313 \\
\bottomrule
\end{tabular}
\caption{Comparing with/without projection to $\widehat{\mathbf{x}}_{0|t}$ in the fidelity step. Direct projection introduces residual texture noise, reducing PSNR and increasing LPIPS.}
\label{tab:avoid_projections}
\vspace{-2mm}
\end{table}



\subsection{Ablations.}
\label{ablation}

\textbf{The effect of $\{ \lambda_t \} $ .} 
We experimentally validate our practical design choices outlined in section \ref{sec:steering the flow}. Having a continuous conditioning with $\lambda = 1$ at every step introduces a noise that is only exacerbated as more steps are taken by the flow model. Our ablations (Figure \ref{fig:lambda_schedule}) verify that explicit conditioning should generally be done at the middle of the schedule. The parameters $(i_{\text{start}},i_{\text{start}},h)$ may be further tuned for better reconstructions.

\textbf{Complex schedules.} $\{ \lambda_i \}$ can be tuned for each task separately, achieving better reconstruction scores. In table \ref{tab:finetuned_schedule} we show the parameters of a two-step window, which has an additional parameter $i_{\text{step}}$, which indicates the step that changes the conditioning strength $\lambda_i$ from $h_1$ to $h_2$. Although more complex schedules can be designed, our results of table\ref{tab:baselines} are based on this two-step schedule.

\begin{figure}[!t]
  \centering
  \includegraphics[width=\textwidth]{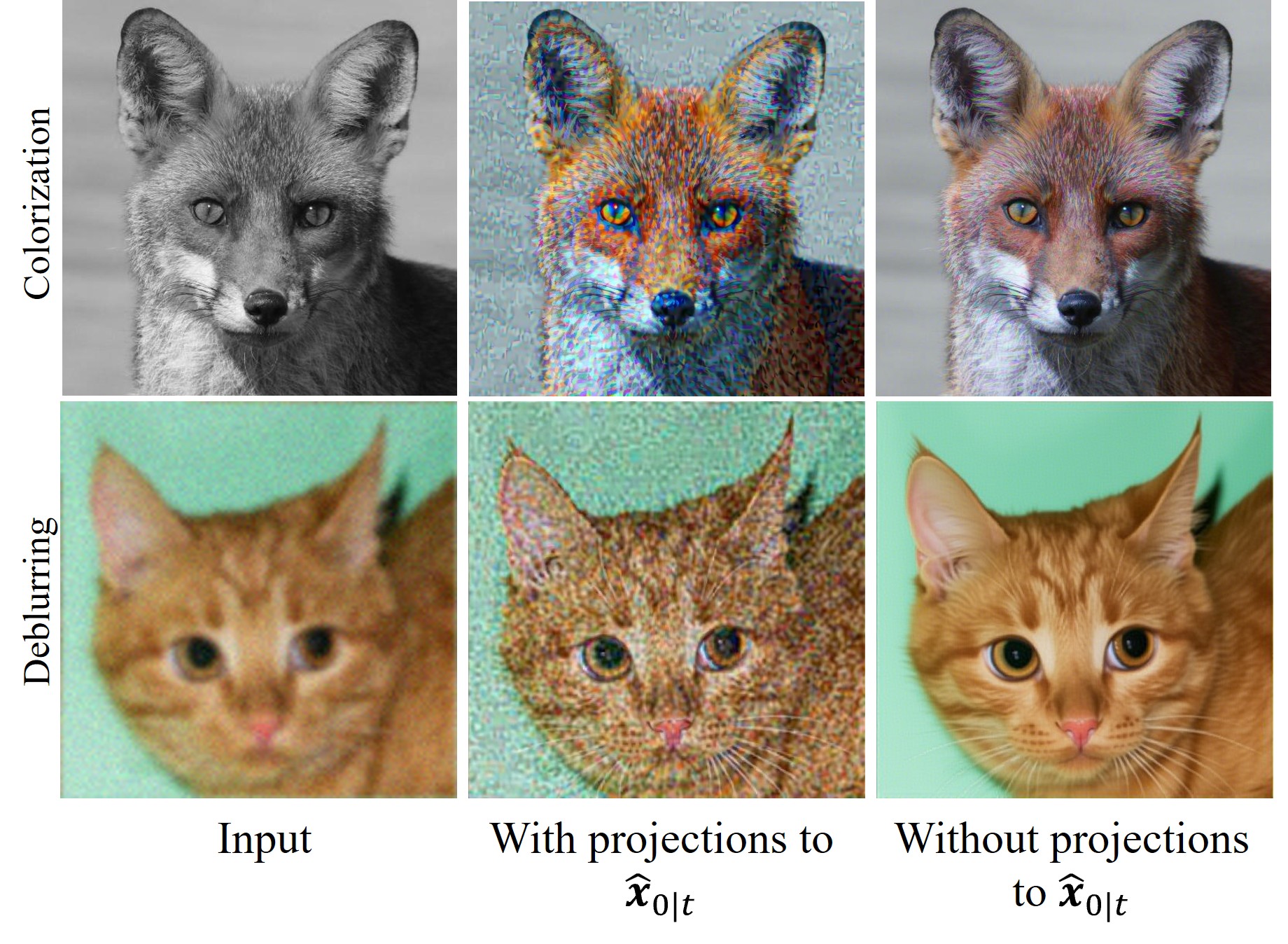}
  \caption{Reconstructed images with and without projecting to $\widehat{\mathbf{x}}_{0|t}$ in the fidelity update step. Projections induce residual noise that persists in the reconstructed images.}
  \label{fig:avoid projections}
\end{figure}
\textbf{Avoiding projections to $\mathbf{\widehat{x}}_{0|t_i}$.}
In Section \ref{sec:steering the flow} we recommend to avoid the linear projection steps $\mathbf{\widehat{x}}_{0|t_i} \leftarrow \mathbf{\widehat{x}}_{t_i}$. This claim is experimentally verified in table \ref{tab:avoid_projections} and figure\ref{fig:avoid projections}. Note the remaining noise of the final reconstructions. The flow model is not trained with a noise schedule $\sigma_t$ in each step. Thus, it conflates noise artifacts with necessary textures, and suggests a velocity update that highlights these unwanted artifacts.

\begin{figure}[!t]
  \centering
  \includegraphics[width=\textwidth]{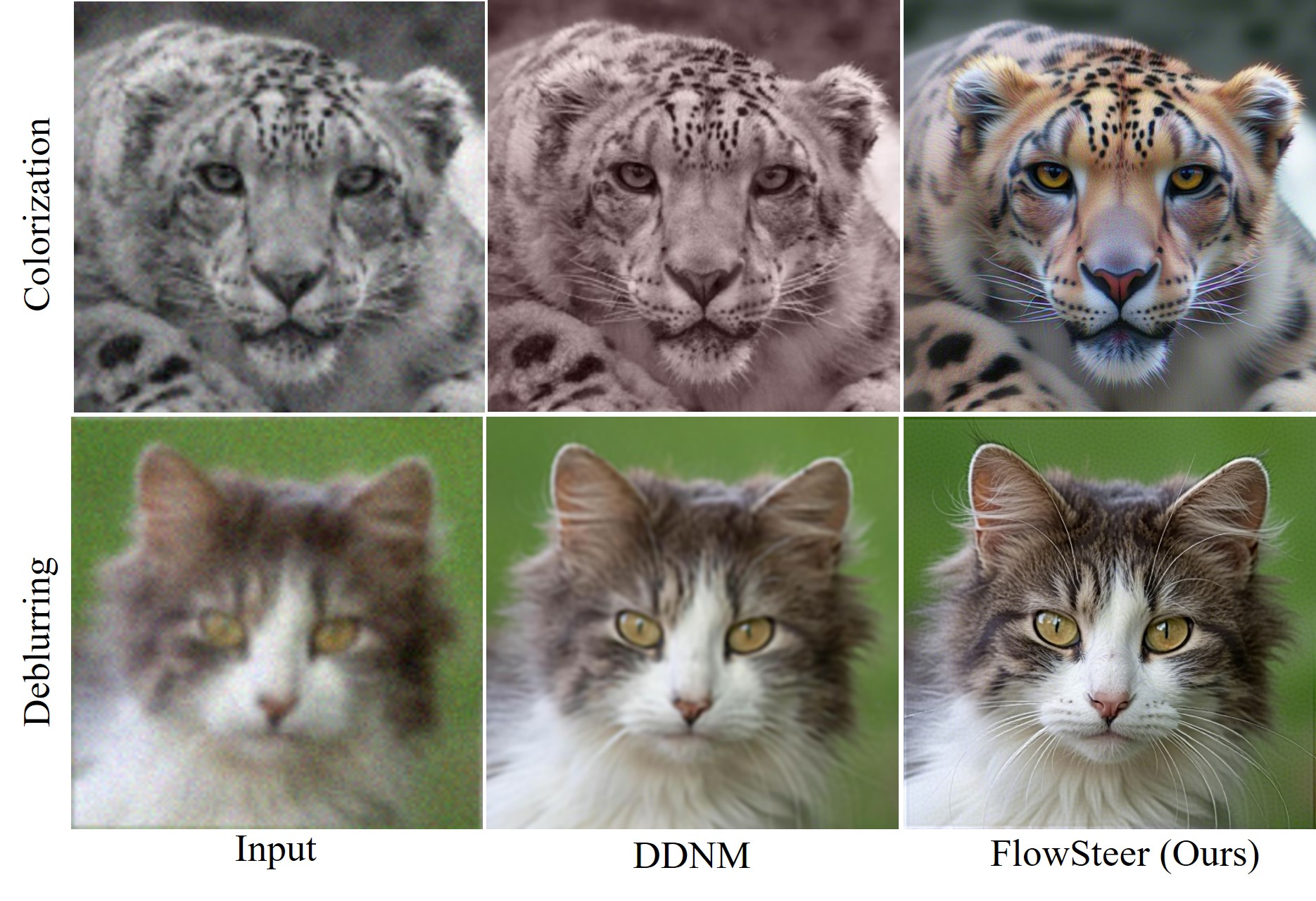}
  \caption{FlowSteer(30 steps) vs DDNM~\cite{wang2022DDNM}, which implements algorithm \ref{alg:ddnm} on guided-diffusion~\cite{openai2021_guideddiffusion} (100 steps).}
  \label{fig:DDNM comparison}
\end{figure}
\textbf{The Diffusion counterpart.} 
We compare Flowsteer with  DDNM~\cite{wang2022DDNM}, which implements algorithm\ref{alg:ddnm} on a pretrained guided-diffusion network~\cite{openai2021_guideddiffusion}. We observe that FlowSteer performs better owing to the powerful generative priors of the flux model. The implementation of the DDNM takes 100 steps on the reconstruction path~\cite{dhariwal2021_diffusionbeatsgans}, while SteerFlow samples the flux model $N=30$ steps for each reconstruction task. 

\section{Conclusion}
Integrating flow-based T2I models into image restoration remains an open problem despite their efficiency over diffusion. FlowSteer articulates this challenge by enabling flow models to better respect the physical forward image formation model. By introducing a simple, yet effective scheduler during the inference process, FlowSteer ensures that the image restoration task will move along a trajectory that will lead to a physically consistent image. More importantly, FlowSteer is training-free, and can be applied to any existing flow-based T2I framework. Across multiple applications, including colorization, super-resolution, deblurring, and denoising, FlowSteer demonstrates both superior pixel-level fidelity and visual quality.
\clearpage
\setcounter{page}{1}
\maketitlesupplementary

\begin{figure*}[!t]
  \centering
  \includegraphics[width=\textwidth]{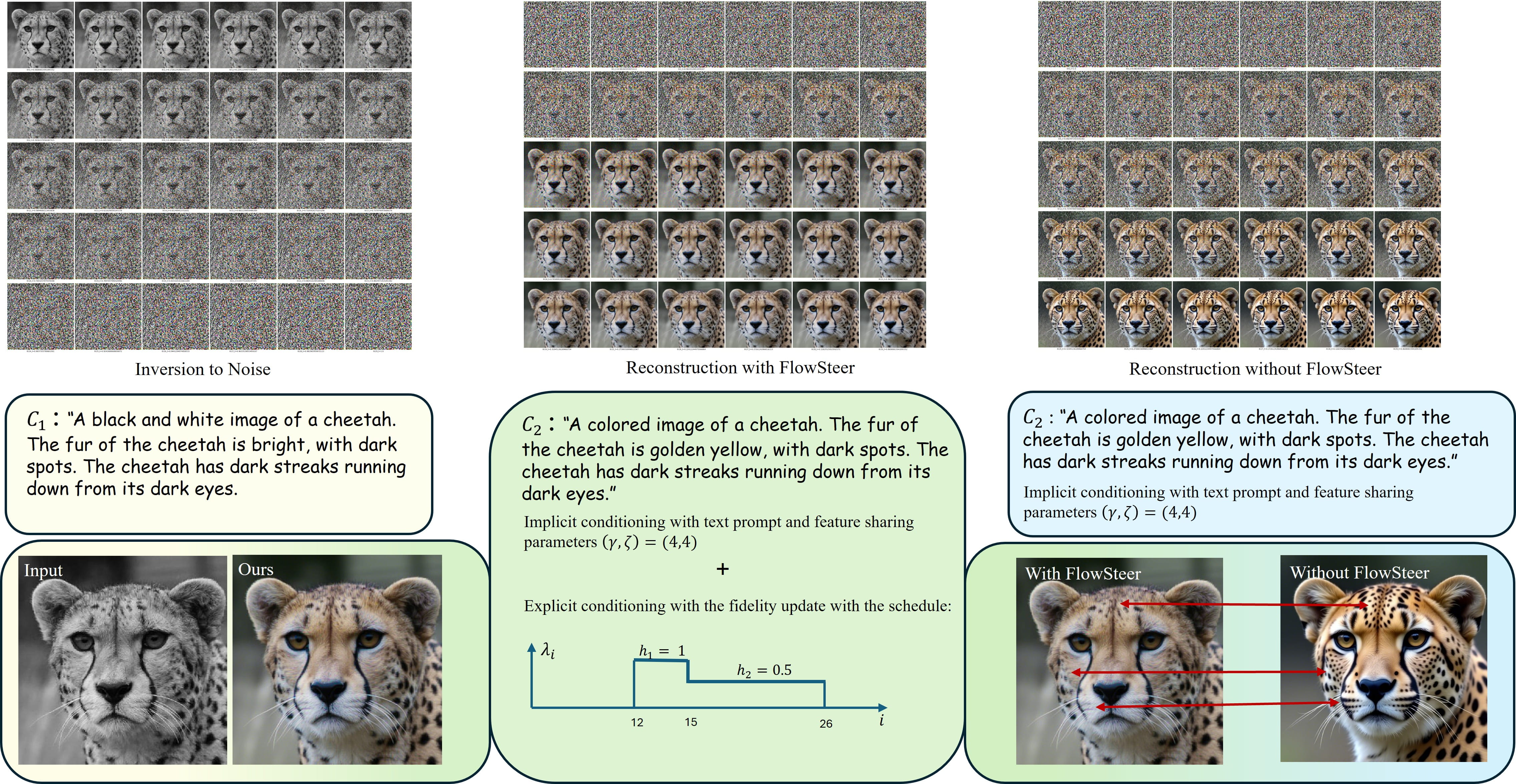}
  \caption{FlowSteer- an overview. Left: The input degraded image is inverted to noise. The source prompt $C_1$ guides the inversion flow path. Middle: The FlowSteer reconstruction path- which is conditioned by the target caption $C_2$, feature sharing parametrized by $(\gamma,\zeta)$, and the fidelity update scheduled by the $\lambda_i$. Right: The FlowSteer output is compared against our baseline flow model without FlowSteer. The implicit conditioning without FlowSteer generates unnecessary hallucinations. (Zoom in for a better view).
        }
  \label{fig:supp_overallframework_withintermediates}
\end{figure*}

\section{Noise Sensitivity of the Flow Model}
\label{sec:robust_noise}

In Section 3, we describe that the flow model is sensitive to noisy intermediate projections. This raises the question: ``Are there noise-robust algorithms from diffusion models, that can be reinterpreted for the flow-model scheme?"
To the best of our knowledge, there are no such algorithms that can be directly translated to the flow scheme. The main reason is that a diffusion model is trained with a noise schedule $\{ \sigma_{t} \}$ for the time schedule $t = {t_N, ... t_0}$. This acts as an inherent buffer of randomness at each step, and is used to design a weight $\lambda_t$ in fidelity update steps. To illustrate this, we present the robust version of Algorithm 1 noise and describe why it cannot be applied directly to the flow model.

\subsection{Noise-robust fidelity update with diffusion}

\paragraph{Fidelity update step.}
Assume a linear forward model with additive noise
\(
\mathbf{y}=\mathbf{A}\mathbf{x}+\boldsymbol{\eta},\quad
\boldsymbol{\eta}\sim\mathcal{N}(\mathbf{0},\sigma_y^2\mathbf{I}).
\)
The DDNM-style fidelity update\cite{wang2022DDNM} for a noisy measurement would update line 4 in Algorithm 1 as follows.
\begin{align}
\widehat{\mathbf{x}}_{0|t}
&= \mathbf{A}^\dagger \mathbf{y} + (\mathbf{I}-\mathbf{A}^\dagger \mathbf{A})\,\mathbf{x}_{0|t} \notag\\
&= \mathbf{A}^\dagger(\mathbf{A}\mathbf{x}+\boldsymbol{\eta})
   +(\mathbf{I}-\mathbf{A}^\dagger \mathbf{A})\,\mathbf{x}_{0|t} \notag\\
&= \mathbf{x}_{0|t}
   - \mathbf{A}^\dagger\big(\mathbf{A}\mathbf{x}_{0|t}-\mathbf{y}\big)
   + \mathbf{A}^\dagger\boldsymbol{\eta}, \quad 
   \boldsymbol{\eta}\!\sim\!\mathcal{N}(\mathbf{0},\sigma_y^2\mathbf{I}),
   \label{eq:ddnm-noisy-end}
\end{align}
which makes explicit the extra noise term \( \mathbf{A}^\dagger\boldsymbol{\eta} \).

\paragraph{Projection back/ Posterior sampling step.}
As in DDPM~\cite{ho2020DDPM} or DDIM~\cite{song2020DDIM}, a diffusion model samples/project back to the reconstruction path at time step $t-1$. This is line 5 of Algorithm 1:
\begin{align}
\mathbf{x}_{t-1} &\sim 
\mathcal{N}\!\Big(\,\boldsymbol{\mu}_t(\mathbf{x}_t,\widehat{\mathbf{x}}_{0|t}),
\,\sigma_t^2\,\mathbf{I}\Big).\notag
\end{align}
With the re-parametrization trick,\\
\begin{align}
\mathbf{x}_{t-1} &= \boldsymbol{\mu}_t(\mathbf{x}_t,\widehat{\mathbf{x}}_{0|t})
   + \sigma_t\,\boldsymbol{\epsilon}_t,\qquad
   \boldsymbol{\epsilon}_t\sim\mathcal{N}(\mathbf{0},\mathbf{I}).
\end{align}
The posterior mean above is
\begin{align}
\boldsymbol{\mu}_t(\mathbf{x}_t,\widehat{\mathbf{x}}_{0|t})
&= 
\underbrace{\frac{\sqrt{\bar{\alpha}_{t-1}}\,\beta_t}{1-\bar{\alpha}_t}}_{=:~a_t}\,
\widehat{\mathbf{x}}_{0|t}
\;+\;
\frac{\sqrt{\alpha_t}\,\big(1-\bar{\alpha}_{t-1}\big)}{1-\bar{\alpha}_t}\,
\mathbf{x}_t .
\label{eq:posterior-mean}
\end{align}

\paragraph{Damped correction($\lambda_t$) and variance matching($\gamma_t$).}
Following the “project-and-correct’’ view, Wang \etal ~\cite{wang2022DDNM} propose to parametrize and dampen the update weight 
\(1 \rightarrow\lambda_t\) for the data-fidelity (null-space) update, and adjust the
diffusion noise by \(\sigma_t \rightarrow \gamma_t\). This changes lines 4 and 5 of Algorithm 1 to the following:
\begin{align}
\widehat{\mathbf{x}}_{0|t}
&= \mathbf{x}_{0|t} - \textcolor{blue}{\lambda_t}\,\mathbf{A}^\dagger\!\big(\mathbf{A}\mathbf{x}_{0|t}-\mathbf{y}\big),
\label{eq:damped-x0}\\
\mathbf{x}_{t-1}
&= \boldsymbol{\mu}_t(\mathbf{x}_t,\widehat{\mathbf{x}}_{0|t})
   + \textcolor{blue}{\gamma_t}\,\boldsymbol{\epsilon}_t,\qquad
   \boldsymbol{\epsilon}_t\sim\mathcal{N}(\mathbf{0},\mathbf{I}). 
\label{eq:damped-sample}
\end{align}
Since \( \widehat{\mathbf{x}}_{0|t} \) contains \( \mathbf{A}^\dagger\boldsymbol{\eta} \),
the sampling mean \eqref{eq:posterior-mean} injects an additional noise
term \(a_t\lambda_t \mathbf{A}^\dagger\boldsymbol{\eta}\) with covariance
\(a_t^2\lambda_t^2\sigma_y^2\,\mathbf{A}^\dagger{\mathbf{A}^\dagger}^{\!*}\).
For the simple linear operators used in the linear tasks selected, they are treated as isotropic
and set \(\gamma_t\) to preserve the target posterior variance. The two principles that guide the adaptive calculation of the new parameters are as follows:\\
i) Variance should be preserved in each step $t$, 
\begin{equation}
\gamma_t^2
\;=\;
\max\!\Big(0,\;\sigma_t^2 \;-\; a_t^2\,\lambda_t^2\,\sigma_y^2\Big). \\
\label{eq:gamma_var-match}
\end{equation}
ii) $\lambda_t$ should be as close as possible to $1$,
\begin{equation}
\lambda_t \;=\;
\begin{cases}
1, & \sigma_t \ge a_t\,\sigma_y,\\[2pt]
\dfrac{\sigma_t}{a_t\,\sigma_y}, & \sigma_t < a_t\,\sigma_y~.
\end{cases}
\label{eq:adaptive_lambda}
\end{equation}

\noindent\textit{Interpretation:}
\(a_t\) is the coefficient on \(\widehat{\mathbf{x}}_{0|t}\) in the posterior mean,
\(\lambda_t\) controls the strength of the fidelity (null-space) correction,
\(\sigma_y\) is the measurement-noise std., and \(\gamma_t\) is the residual diffusion noise after accounting for injected measurement noise.

\subsection{Can this scheme extend to flow models?}
Reframe equation \ref{eq:gamma_var-match} as $\gamma_t^{2} \;=\; \sigma_t^{2} \;-\; \big(a_t\,\lambda_t\,\sigma_y\big)^{2}$ leads to the following condition for $\gamma_t$:
\begin{equation}
\gamma_t \;=\;
\begin{cases}
\sigma_t^{2} \;-\; \big(a_t\,\sigma_y\big)^{2}, & \sigma_t \ge a_t\,\sigma_y,\\[2pt]
0, & \sigma_t < a_t\,\sigma_y~.
\end{cases}
\label{eq:adaptive_gamma}
\end{equation}

Since flow models do not have a noise schedule $\sigma_t$, it is equivalent to setting $\sigma_t = 0$. This would imply the second case $(\sigma_t < a_t\,\sigma_y)$ of equations \ref{eq:adaptive_lambda} and \ref{eq:adaptive_gamma}, leading to $\gamma_t = 0$ and
\begin{equation}
    \lambda_t = \dfrac{\sigma_t}{a_t\,\sigma_y} = 0.
\end{equation}

Thus, in a flow model with no noise schedule $\sigma_t \equiv 0$ the noise-aware formulation collapses to $\gamma_t = 0$ and $\lambda_t = 0$—i.e., no fidelity correction. Moreover, the core principle of ``preserving the variance in each timestep $t$" does not apply, since a flow model does not have an inherent variance schedule through $\sigma_t$. This shows that the diffusion-style noise-robust update does not transfer directly to flows. Empirically, we find that fidelity updates in flow models are sensitive to measurement noise and, if used at all, must be applied sparingly with careful, task-dependent tuning.


\section{More Details on the Flow Model}
\label{sec:network}
FlowSteer, and all related baselines were implemented on an NVIDIA A100 GPU with VRAM 80GB. Resources were only required for model inference.

\subsection{Feature sharing}
The Flux-dev~\cite{FLUX2024} model is our base-line image editing model, that we adapt for image restoration. Both the inversion path and the reconstruction path has $N=30$ steps. The velocity prediction $v_{\theta}(.)$ in each step is modeled through a Diffusion Transformer (DiT)~\cite{esser2024_DiT} block. The diffusion transformer has ``double-block" layers and ``single block" layers, out of which the ``single block" layers are used for feature sharing. The attention maps that are calculated in the last $\zeta$ inversion steps (input to noise) and correspond to the first $\zeta$ reconstruction steps (noise to image). During reconstruction, the cached attention maps are used for the first $\zeta$ steps, creating an implicit conditioning that the reconstructed image should preserve some qualities of the original image. There are similar approaches in literature of caching attention maps or just the Values or Key-Query pairs to drive an edited image/video to be faithful to an input image.~\cite{hertz2022_prompttoprompt, cao2023masactrl,hertz2024style,zhang2023real,wang2024_RFEdit, kim2025reflex_textguided, geyer2023tokenflow, qi2023fatezero, liu2024_videop2p,ceylan2023pix2video}. In our design, we cache the complete attention map. However, after experimenting with different feature sharing schemes and grid searches through hyper-parameters as described in section 3, we find that such implicit conditioning is insufficient for enforcing the pixel-level fidelity required for image restoration tasks.

\subsection{Text prompts}
As described in section 3, there are two types of prompts that implicitly condition the flow field. During inversion, the source prompt $C_1$ is used to describe the degraded image. During reconstruction, the target prompt $C_2$ is used to describe the target image features. The prompts are manually selected by the user. The table \ref{tab:prompt_pairs} shows the sample prompts used for each of the restoration tasks.

\begin{table*}[t]
\centering
\small
\setlength{\tabcolsep}{4pt}
\renewcommand{\arraystretch}{1.15}
\begin{tabularx}{\textwidth}{@{} l c Y Y @{}}
\toprule
\textbf{Type} & \textbf{\#} & \textbf{Source prompt }$C_1$ & \textbf{Target prompt }$C_2$ \\
\midrule

\multicolumn{4}{@{}l}{\textit{Colorization}}\\[2pt]
Pets & 1 &
A black and white image of a cat. The cat is white with black patches. The background is dark. The nose of the cat is pink. The eyes of the cat are green. &
A colorful image of a cat. The cat is white with black patches. The background is dark. The nose of the cat is pink. The eyes of the cat are green. \\
Wild & 2 &
A black and white image of a cheetah. The fur of the leopard is bright, with dark spots. The leopard has dark streaks running down from its dark eyes. &
A colored image of a cheetah. The fur of the leopard is golden yellow, with dark spots. The leopard has dark streaks running down from its dark eyes. \\
Wild & 3 &
A black and white image of a fox. The fox has brown fur with white streaks. The nose of the fox is black. The eyes of the fox are dark brown. &
A colorful image of a fox. The fox has brown fur with white streaks. The nose of the fox is black. The eyes of the fox are dark brown. \\
Humans & 4 &
A black and white image of a man. &
A colorful image of a man. The man has black eyes. \\

\addlinespace[2pt]
\multicolumn{4}{@{}l}{\textit{Deblurring}}\\[2pt]
Pets & 1 & A blurred image of a cat. The cat has white fur with black spots. & A sharp image of a cat. The cat has white fur with black spots. Highly detailed, taken using a Canon EOS R camera, hyper detailed photo-realistic maximum detail. \\
Wild & 2 & A blurred image of a lion. The lion has golden colored fur and dark brown eyes. & A sharp image of a lion. The lion has golden colored fur and dark brown eyes. Highly detailed, taken using a Canon EOS R camera, hyper detailed photo-realistic maximum detail. \\
Wild & 3 & A blurred image of a leopard. The leopard has brown fur and black patches. & A sharp image of a leopard. The leopard has brown fur and black patches. Highly detailed, taken using a Canon EOS R camera, hyper detailed photo-realistic maximum detail. \\
Humans & 4 & A blurred image of a man. & A sharp image of a man. Highly detailed, taken using a Canon EOS R camera, hyper detailed photo-realistic maximum detail. \\

\addlinespace[2pt]
\multicolumn{4}{@{}l}{\textit{Super-resolution}}\\[2pt]
Pets & 1 & A low resolution image of a dog. The dog is brown with white patches. & A sharp, high resolution image of a dog. The dog is brown with white patches. Highly detailed, taken using a Canon EOS R camera, hyper detailed photo-realistic maximum detail. \\
Wild & 2 & A low resolution image of a lion. The lion has golden colored fur and dark brown eyes. & A sharp, high resolution image of a lion. The lion has golden colored fur and dark brown eyes. Highly detailed, taken using a Canon EOS R camera, hyper detailed photo-realistic maximum detail. \\
Humans & 3 & A low resolution image of a woman. & A sharp, high resolution image of a woman. Highly detailed, taken using a Canon EOS R camera, hyper detailed photo-realistic maximum detail. \\

\addlinespace[2pt]
\multicolumn{4}{@{}l}{\textit{Denoising}}\\[2pt]
Pets & 1 & A noisy image of a dog. The dog is brown with white patches. & A clean, noise free image of a dog. The dog is brown with white patches. Highly detailed, taken using a Canon EOS R camera, hyper detailed photo-realistic maximum detail. There are no RGB noise artifacts. \\
Wild & 2 & A noisy image of a cheetah. The fur of the cheetah is bright, with dark spots. The cheetah has dark streaks running down from its dark eyes. & A clean, noise free image of a cheetah. The fur of the cheetah is bright, with dark spots. The cheetah has dark streaks running down from its dark eyes. Highly detailed, taken using a Canon EOS R camera, hyper detailed photo-realistic maximum detail. There are no RGB noise artifacts. \\
Human & 3 & A noisy image of a man. & A clean, noise free image of a man. Highly detailed, taken using a Canon EOS R camera, hyper detailed photo-realistic maximum detail. There are no RGB noise artifacts. \\

\bottomrule
\end{tabularx}
\caption{Sample prompt pairs per task. “Source” describes the input (e.g., grayscale, low-res, noisy, blurred); “Target” describes the intended restored image that is used to implicitly steer the flux model.}
\label{tab:prompt_pairs}
\end{table*}

\section{Details on the Degradation Models}
\label{sec:degradation_models}

\textbf{Colorization.}
The forward operator $\mathbf{A}$ maps RGB to an achromatic image by averaging
channels and repeating the result across three channels. For
$\mathbf{x}\in\mathbb{R}^{3\times H\times W}$ and pixel $p$,
\[
(\mathbf{A}\mathbf{x})_c(p)
= \tfrac{1}{3} \ \sum_{k=1}^{3} x_k(p),\qquad c\in\{1,2,3\}.
\]
The Moore–Penrose pseudoinverse coincides with $\mathbf{A}$, i.e.
\[
\mathbf{A}^{\dagger}\mathbf{y}
=\mathcal{R}\ \left( \tfrac{1}{3}\ \sum_{k=1}^{3} y_k\right),
\]
where $\mathcal{R}(\cdot)$ replicates a single channel three times.
Hence, for any $\mathbf{y}\in\mathrm{range}(\mathbf{A})$ (three identical channels),
$\mathbf{A}^{\dagger}\mathbf{y}=\mathbf{y}$ and $\mathbf{A}\mathbf{A}^{\dagger}=\mathbf{A}$.

\paragraph{Deblurring.} The forward operator is a circular Gaussian blur
$\mathbf{A}$ (convolution with kernel $h$). We use the Tikhonov-regularized
(Wiener-type) pseudoinverse parameterized by $\lambda_{\text{W}}$,
\[
\mathbf{A}^{\dagger}_{\lambda_{\text{W}}} \;\triangleq\; (\mathbf{A}^{\!*}\mathbf{A}+\lambda_{\text{W}} \mathbf{I})^{-1}\mathbf{A}^{\!*},
\]
where $\mathbf{A}^{\!*}$ denotes the adjoint of $\mathbf{A}$. We set $\lambda_{\text{W}} = 0.1$ in all our experiments. The forward model can be implemented using Fourier transforms as $\mathcal{F}\{\mathbf{y}\} = \mathcal{F}\{h\}\,\mathcal{F}\{\mathbf{x}\}$, (or equivalently, as the convolution $\mathbf{y}=h * \mathbf{x}$).
The inversion is implemented in python code in the following form,
\[
\mathbf{A}^{\dagger}_{\lambda_{\text{W}}}\mathbf{y}
= \mathcal{F}^{-1}\!\left[\frac{\overline{\mathcal{F}\{ h\}}}{|\mathcal{F}\{ h\}|^{2}+\lambda_{\text{W}}}\,\mathcal{F}\{\mathbf{y}\}\right].
\]
$\overline{\mathcal{F}\{ h\}}$ is the element-wise complex conjugate of $\mathcal{F}\{ h\}$. For a real, symmetric Gaussian kernel $h$, $\overline{\mathcal{F}\{ h\}}=\mathcal{F}\{ h\}$.

\paragraph{Super-resolution ($\times 4$).}
The forward operator $\mathbf{A}$ downsamples an RGB image by average–pooling over non-overlapping $4\times4$ blocks (per channel) and decimating by a factor of $4$ along height and width:
$\mathbf{y}=\mathbf{A}\mathbf{x}\in\mathbb{R}^{3\times H/4\times W/4}$ with
\[
y_c(u,v)=\tfrac{1}{16}\!\!\sum_{i,j=0}^{3} x_c(4u+i,\,4v+j).
\]
As a practical pseudo-inverse we use a right-inverse that restores the original spatial size by \emph{patch replication} (nearest-neighbor up-sampling):
\[
(\mathbf{A}^{\dagger}\mathbf{y})_c(i,j)
=\mathbf{y}_c\!\big(\!\lfloor i/4\rfloor,\,\lfloor j/4\rfloor\!\big),
\]
such that $\mathbf{A}\mathbf{A}^{\dagger}=\mathbf{I}\text{ on }\mathbb{R}^{3\times H/4\times W/4}$.

\paragraph{Denoising.}
We model denoising with the identity forward operator
$\mathbf{A}=\mathbf{I}$, so the measurement is simply
$\mathbf{y}=\mathbf{A}\mathbf{x}+\boldsymbol{\eta}=\mathbf{x}+\boldsymbol{\eta}$.
Because $\mathbf{I}$ is self-adjoint and full-rank, its Moore–Penrose pseudoinverse is itself: $\mathbf{A}^{\dagger}=\mathbf{I}$.

\section{Design Details for \texorpdfstring{$\{\lambda_i\}$}{lambda_i schedules}}
\label{sec:lambda_schedules}

\subsection{A Two-Step Schedule}
In section 5.3 we describe the effect of the schedule $\{ \lambda_i \}$. Empirically we observe that a single step-design as mentioned in table 2 (left) is sufficient for general reconstructions. However, the reconstruction quality can be improved by having a scheduler that gradually reduces the strength of the fidelity update. This empirical observation is shown in table 2 (right).\\ 

The following is the python-style implementation of a two-step scheduler, with the parameters $i_{\text{start}}, i_{\text{step}}, i_{\text{end}} , h_1, h_2$. Between $i_{\text{start}}, i_{\text{step}}$ the value of $\lambda_i$ is $h_1$, and between $i_{\text{step}}, i_{\text{end}}$ the value of $\lambda_i$ is $h_2$. 

\begin{lstlisting}[style=python-ddnm,
  caption={Step-shaped $\lambda_i$ schedule used in SteerFlow.}]
def make_lambda_step_schedule(
    timesteps,
    *, start, step, end,
    h_1=1.0, h_2=0.5,
    final_pad=1,
):
    N = len(timesteps) - 1
    lam = np.zeros(N, dtype=np.float32)
    if N <= 0:
        return lam

    i_0 = to_index(start, N)
    i_1 = to_index(step,  N)
    # make end exclusive
    i_2 = to_index(end,   N) + 1   

    # order & clamp
    i_start, i_step = min(i_0, i1), max(i0, i1)
    i_step, i_end = min(i1, i2), max(i1, i2)
    i_start = np.clip(i_start, 0, N)
    i_step = np.clip(i_step, 0, N)
    i_end = np.clip(i_end, 0, N)

    if i_start < i_step:
        lam[i_start:i_step] = h_1
    if i_step < i_end:
        lam[i_step:i_end] = h_2

    if 0 < final_pad < N:
        lam[-final_pad:] = 0.0

    # keep a peak=1 if padding nuked it
    if lam.max() <= 0 and N - final_pad - 1 >= 0:
        lam[N - final_pad - 1] = 1.0
    else:
        lam /= max(1.0, float(lam.max()))
    return lam
\end{lstlisting}

\subsection{Effect of the Fidelity update}
\begin{figure*}[!t]
  \centering
  \includegraphics[width=\textwidth]{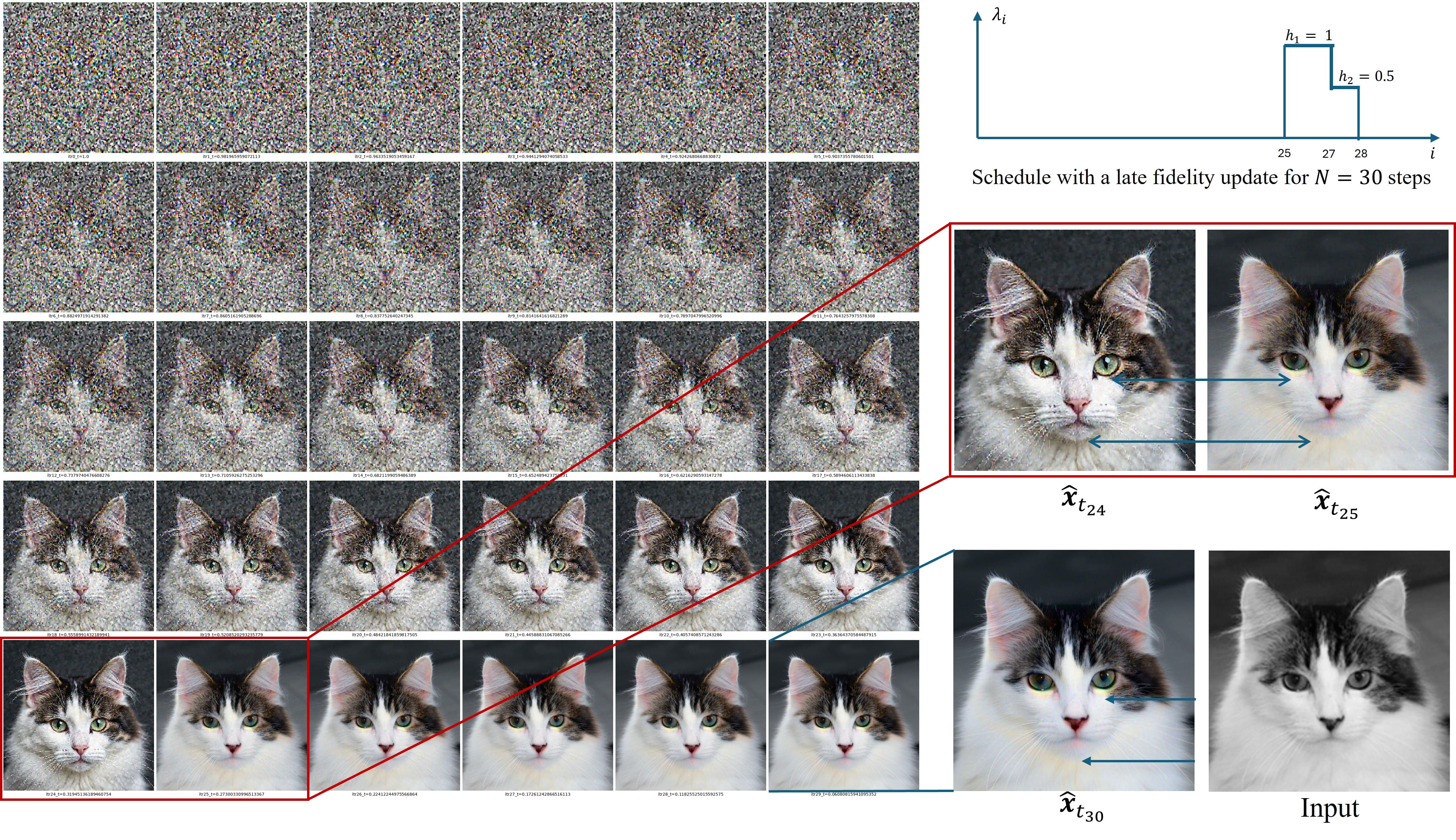}
  \caption{Visualising the reconstruction path with the fidelity  update being late. Before the fidelity update has been appled ($\mathbf{\widehat{x}}_{t_{24}}$), the flux model has already hallucinated fine textures and colors. Immediately after activation ($\mathbf{\widehat{x}}_{t_{25}}$), the fidelity update steers the path toward the measurement. However, some residual artifacts persist in the final reconstruction.}
  \label{fig:supp_latefidelity}
\end{figure*}

\begin{figure*}[!t]
  \centering
  \includegraphics[width=\textwidth]{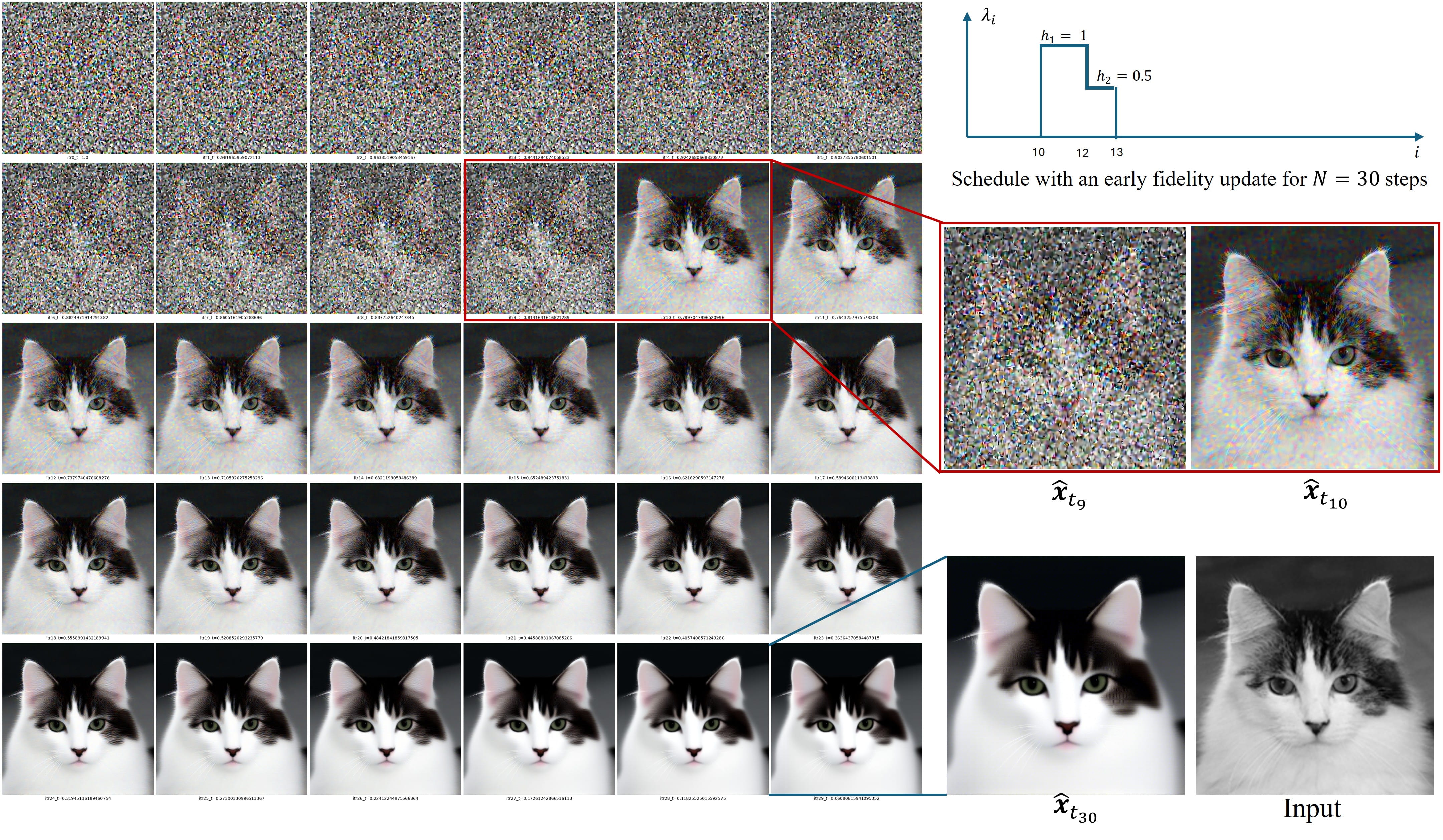}
  \caption{Visualising the reconstruction path with the fidelity update being early. Before the fidelity update has been appled ($\mathbf{\widehat{x}}_{t_{9}}$), the flux model has not completely formed the color pallette and textures. The fidelity update conditions the path to have less color($\mathbf{\widehat{x}}_{t_{10}}$), and the large number of steps without the fidelity update over smoothens the result. The same effect will be seen when a higher number of flux steps are padded at the end of the schedule, as described in Section \ref{sec:supp_finalpadding}.}
  \label{fig:supp_earlyfidelity}
\end{figure*}

The effect of the fidelity update is visually shown in the figures \ref{fig:supp_latefidelity} and \ref{fig:supp_earlyfidelity}. If the update is started too late ($i_{\text{start}}$ is too late), then the reconstruction will have artifacts from the hallucinated details. (figure \ref{fig:supp_latefidelity}). If the update is started too early($i_{\text{start}}$ is too early), then the desired level of hallucinations (such as rich colors for colorization) have not yet formed. This makes the reconstruction have over smoothness, and loose color and texture details. (figure \ref{fig:supp_earlyfidelity}).

\subsection{Final padding}
\label{sec:supp_finalpadding}
A final padding parameter is added to enforce that at least a few steps of the reconstruction path does not end with a fidelity update. This is another empirical design decision, where we observe that having some steps without fidelity updates helps to smoothen out some of the noise artifacts. However, if the final padding is too high, the image can get oversmoothed. This is effectively the same as having a fidelity update too early on in the reconstruction path, as shown in figure \ref{fig:supp_earlyfidelity}.

\section{Limitations and Future Directions}
\label{sec:limitations}

\subsection{Empirical schedule}
\begin{figure*}[!t]
  \centering
  \includegraphics[width=\textwidth]{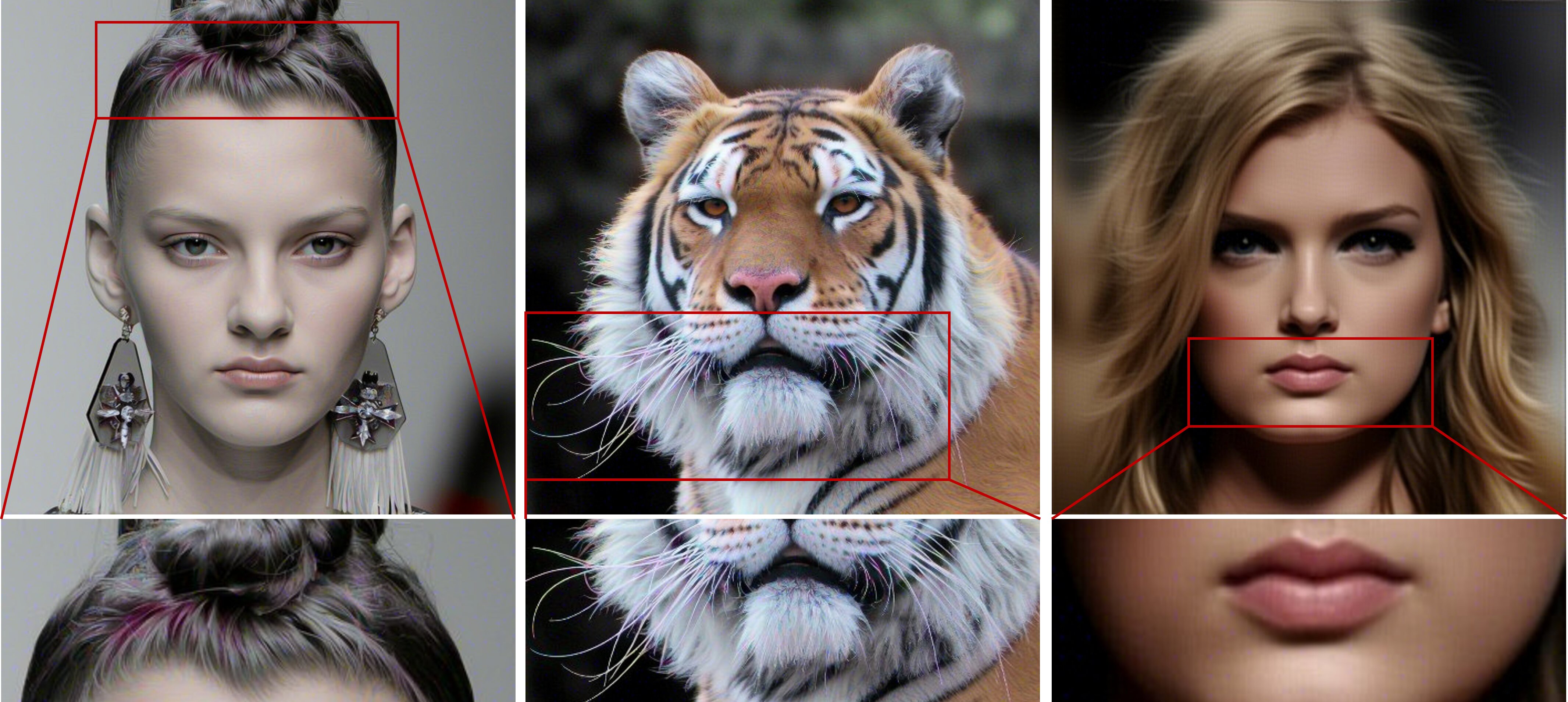}
  \caption{Noise artifacts are seen in some of the final reconstructed image with the two-step scheduler. The restoration tasks corresponding to these images are colorization (left and center) and deblurring (right). Fine tuning the scheduler on a per-image basis may reduce these artifacts as discussed in Section \ref{sec:limitations}. Zoom in to clearly see the noise artifacts in some pixels.}
  \label{fig:supp_limitations}
\end{figure*}

The Flow Steer schedule recommendations are given through empirical fine-tuning. For the four restoration tasks together, we recommend a one-step schedule in Table 2(left). For each task separately, we recommend a two-step schedule in Table 2(right). Even with this task-specific schedule recommendation, there are cases where the schedule does not work for every image in the dataset. This can be seen in some of the failure cases for the current SteerFlow scheme as shown in Figure \ref{fig:supp_limitations}. 

A future directions would be to extend this schedule to be adaptable for each individual image, so that manual tuning can be avoided. The key is to estimate the noise injected by the fidelity update and trigger the update at the step whose model noise budget best matches it, while adjusting the update strength $\{ \lambda_t \}$ accordingly. This would lead to an image-adaptive scheduler, and will not have to depend on heuristics.

\subsection{Artifacts from explicit conditioning}
\begin{figure*}[!t]
  \centering
  \includegraphics[width=\textwidth]{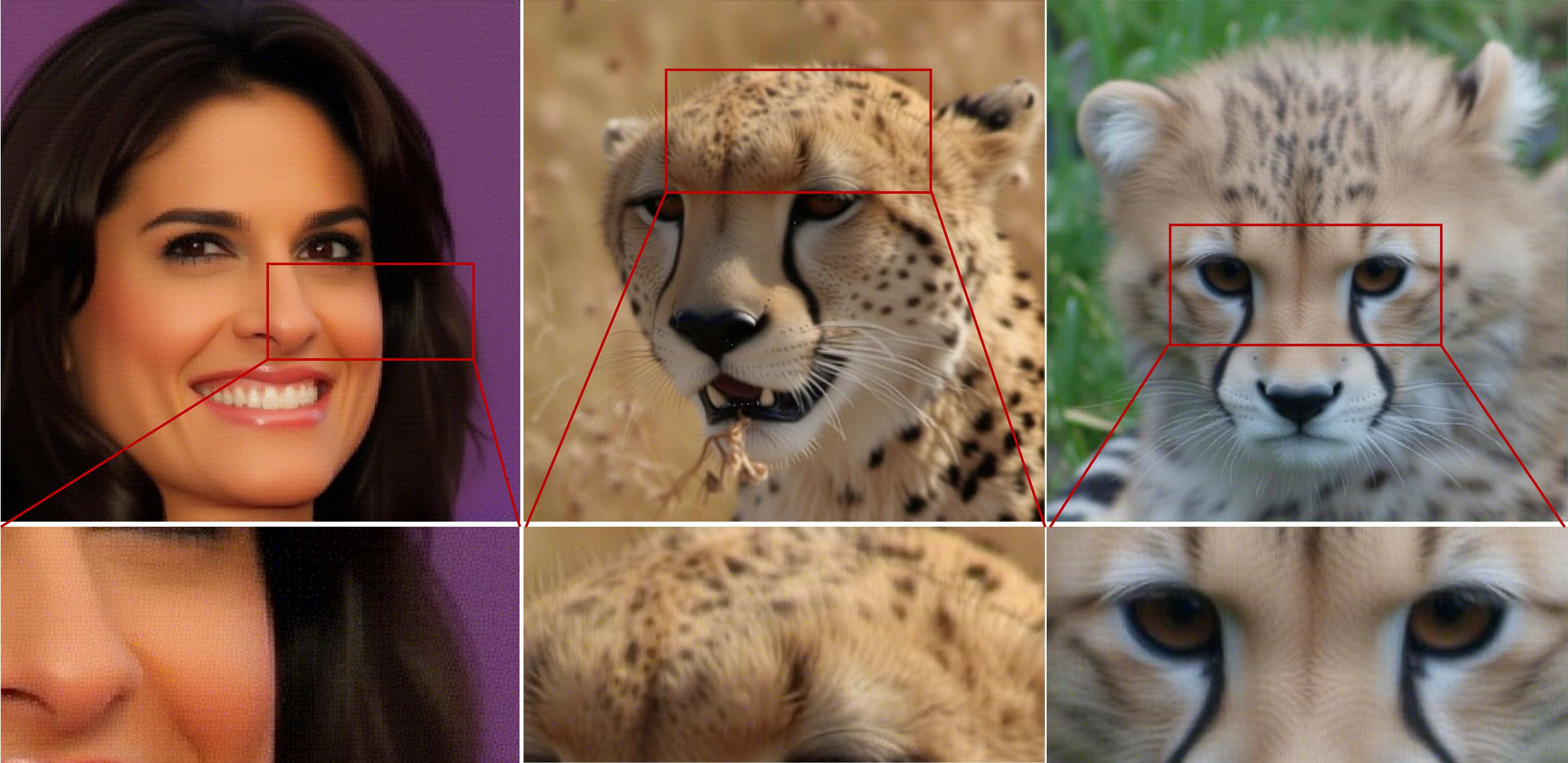}
  \caption{ Artifacts resulting from Pseudo-inverse operators in FlowSteer as discussed in Section \ref{sec:limitations}. Blocking artifacts affect edges in super-resolution(left). The Weiner filter creates ringing effects in deblurring(center and right). Zoom in to clearly see the artifacts.}
  \label{fig:supp_pinv_limitations}
\end{figure*}

The explicit conditioning of FlowSteer relies on Pseudoinverse operators. This has the inherent limitation of introducing artifacts, such as shown in figure \ref{fig:supp_pinv_limitations}. Apart from the noisy artifacts described above, there are instances of blocking-artifacts from the upsampling operation in super-resolution, and ringing-artifacts from the Weiner filter in deblurring.

\section{More Visual Results}
\label{sec:visual_results}

\subsection{More visual results on restoration tasks}
More images on the restoration tasks, highlighting that we preserve both pixel-level fidelity and rich perceptual quality. Zoom in and highlight differences.

\subsection{Plug-and-Play(PnP) on pre-trained Flow models}
\begin{figure*}[!t]
  \centering
  \includegraphics[width=\textwidth]{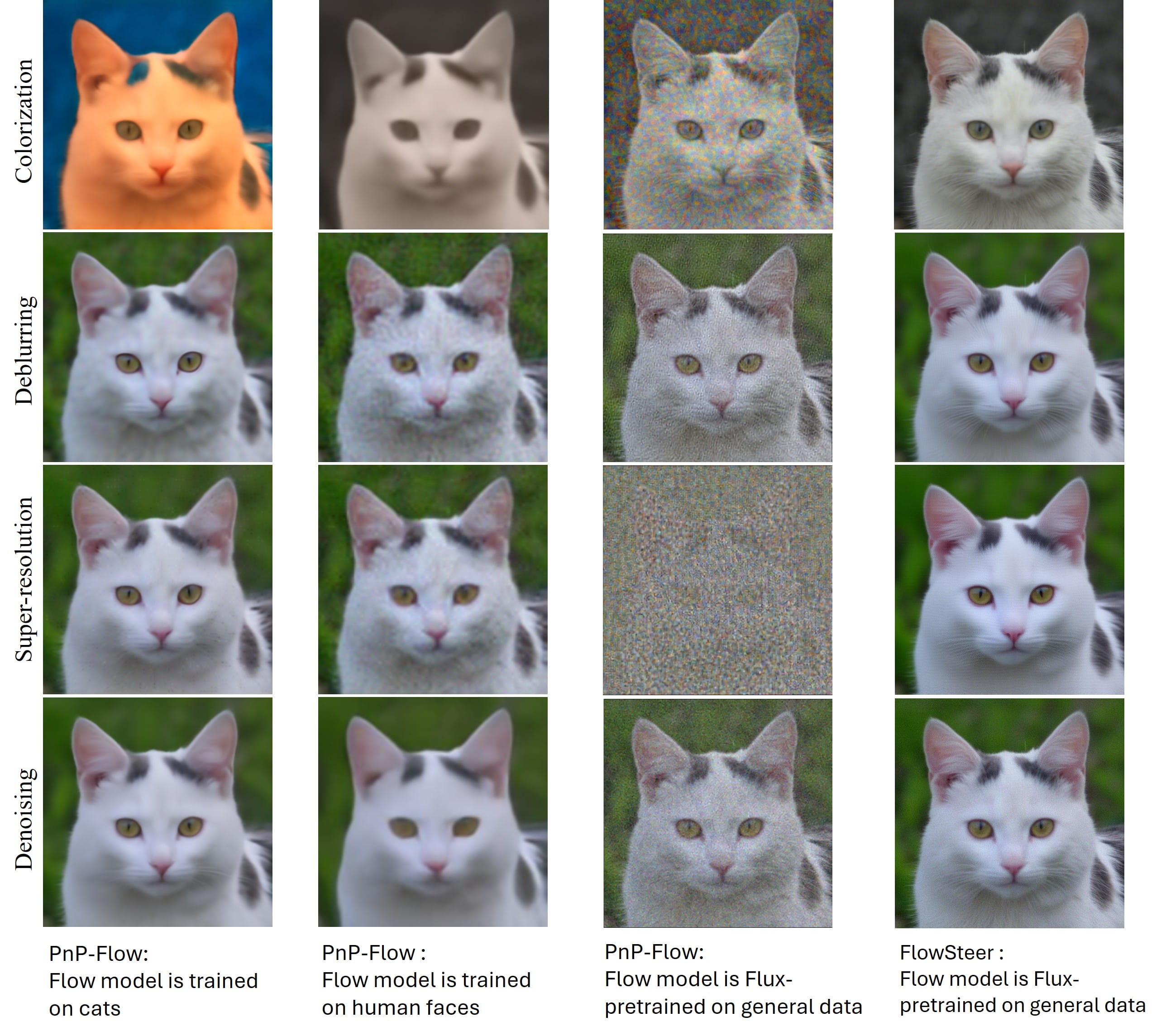}
  \caption{ Plug and play approach on different types of flow models. Although it produces plausible results on a flow model trained on cats(left), it does poorly when implemented on a flow model trained on other general data. This is described in Section \ref{sec:visual_results}.}
  \label{fig:supp_pnp_flow_versions}
\end{figure*}

We recreate the results of PnP-Flow~\cite{martin2024pnpflow} to verify if a Plug-and-play type algorithm can achieve visually appealing images after image restoration. It was observed that it performs well only when the underlying model is the provided model from the authors, which is specifically trained for the class of images being tested. For example, with the pre-trained flow model (which has been trained on a cat dataset) it gives plausible results on cat images. However, when flux-dev~\cite{FLUX2024} (which is trained on a much broader class of data) is used as the underlying flow model, the Plug-and-play method fails to converge to a plausible reconstruction. A grid search was run to select the hyper-parameters of the PnP algorithm and the values: $(\alpha, \gamma, \eta_{dn}) = (0.3,0.8,0.3)$ were selected. This is demonstrated in Figure \ref{fig:supp_pnp_flow_versions}.

\begin{figure*}
\centering
\begingroup
\setlength{\tabcolsep}{0pt} 
\renewcommand{\arraystretch}{0.1} 
\begin{tabular}{cccccccc} 
    
    \multirow{2}{*}[4ex]{\rotatebox[origin=c]{90}{\textbf{Colorization}}} &
    \includegraphics[width=0.140\textwidth, angle=0, clip]{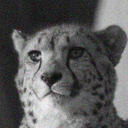} & 
    \includegraphics[width=0.140\textwidth, angle=0, clip]{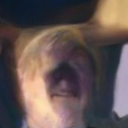} &
    \includegraphics[width=0.140\textwidth, angle=0, clip]{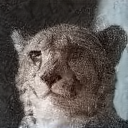} &
    \includegraphics[width=0.140\textwidth, angle=0, clip]{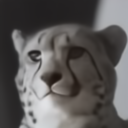} &
    \includegraphics[width=0.140\textwidth, angle=0, clip]{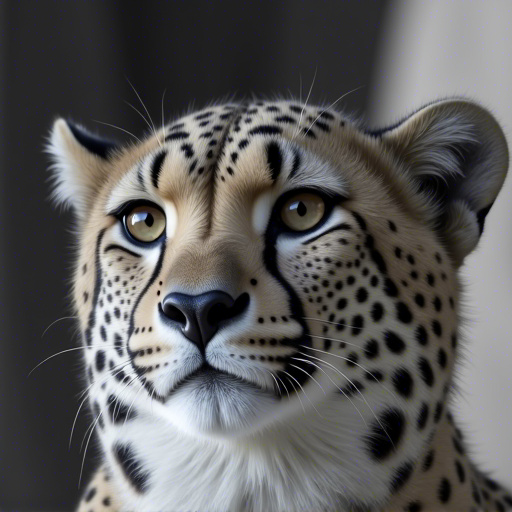} &
    \includegraphics[width=0.140\textwidth, angle=0, clip]{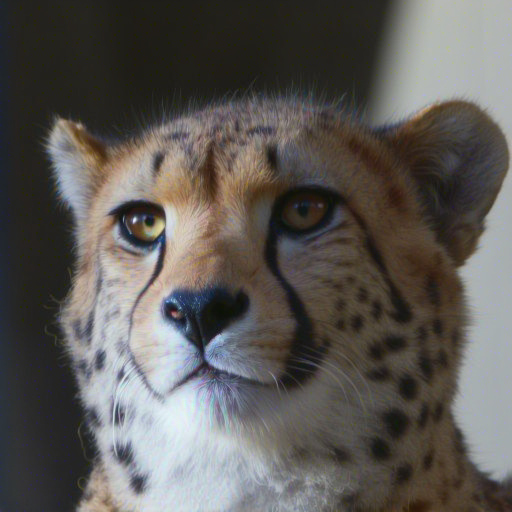} &
    \includegraphics[width=0.140\textwidth, angle=0, clip]{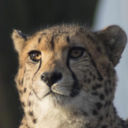}\\
    
    &
    \includegraphics[width=0.140\textwidth, angle=0, clip]{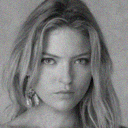} & 
    \includegraphics[width=0.140\textwidth, angle=0, clip]{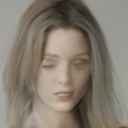} &
    \includegraphics[width=0.140\textwidth, angle=0, clip]{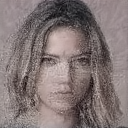} &
    \includegraphics[width=0.140\textwidth, angle=0, clip]{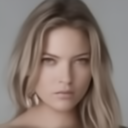} &
    \includegraphics[width=0.140\textwidth, angle=0, clip]{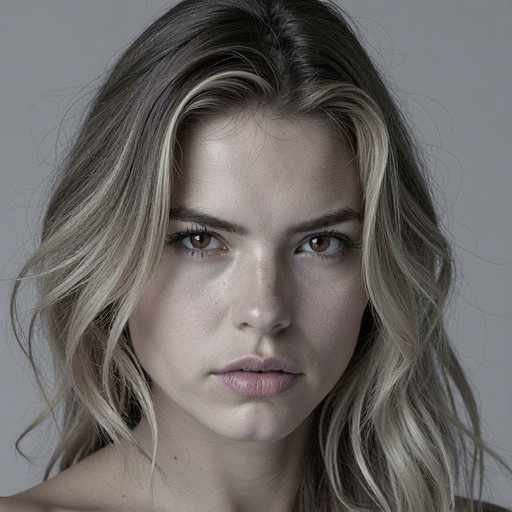} &
    \includegraphics[width=0.140\textwidth, angle=0, clip]{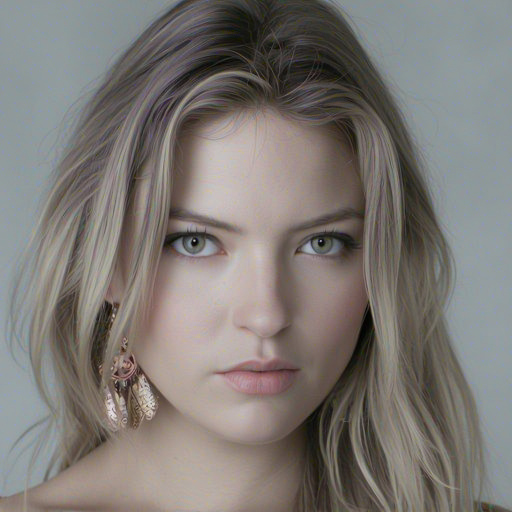} &
    \includegraphics[width=0.140\textwidth, angle=0, clip]{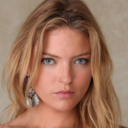} \\[0.1cm]
    
    \multirow{2}{*}[8ex]{\rotatebox{90}{\textbf{Super-resolution}}} &
    \includegraphics[width=0.140\textwidth, angle=0, clip]{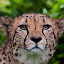} & 
    \includegraphics[width=0.140\textwidth, angle=0, clip]{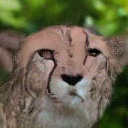} &
    \includegraphics[width=0.140\textwidth, angle=0, clip]{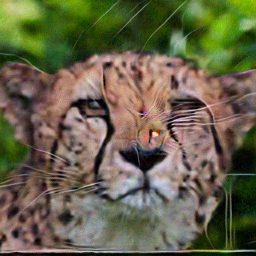} &
    \includegraphics[width=0.140\textwidth, angle=0, clip]{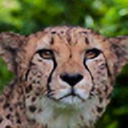} &
    \includegraphics[width=0.140\textwidth, angle=0, clip]{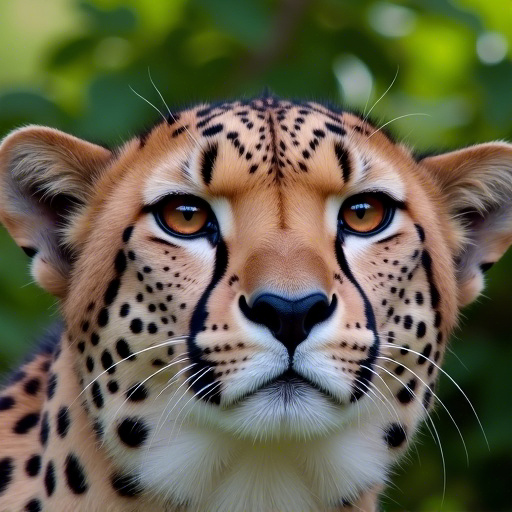} &
    \includegraphics[width=0.140\textwidth, angle=0, clip]{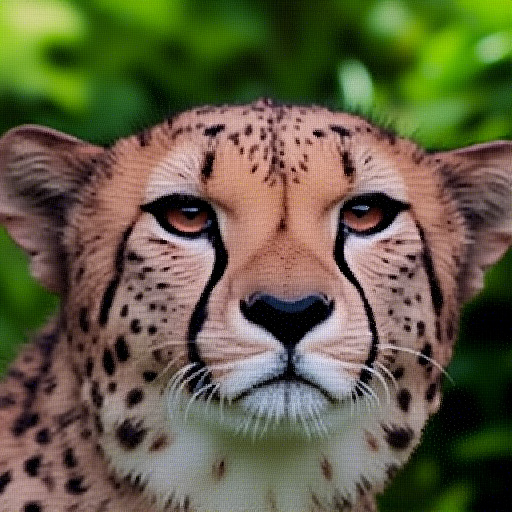} &
    \includegraphics[width=0.140\textwidth, angle=0, clip]{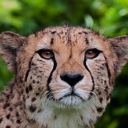} \\

    &
    \includegraphics[width=0.140\textwidth, angle=0, clip]{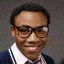} & 
    \includegraphics[width=0.140\textwidth, angle=0, clip]{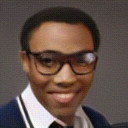} &
    \includegraphics[width=0.140\textwidth, angle=0, clip]{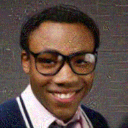} &
    \includegraphics[width=0.140\textwidth, angle=0, clip]{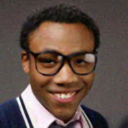} &
    \includegraphics[width=0.140\textwidth, angle=0, clip]{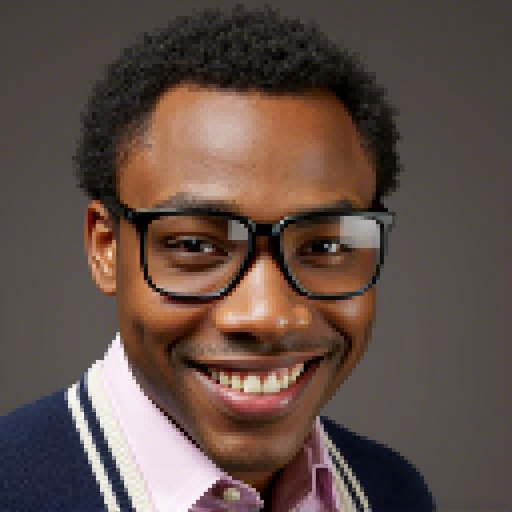} &
    \includegraphics[width=0.140\textwidth, angle=0, clip]{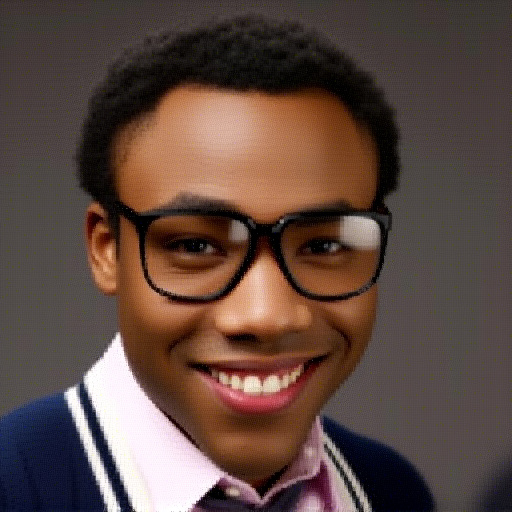} &
    \includegraphics[width=0.140\textwidth, angle=0, clip]{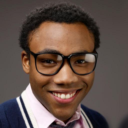} \\[0.1cm]
    
    \multirow{2}{*}[4ex]{\rotatebox{90}{\textbf{Denoising}}} &
    \includegraphics[width=0.140\textwidth, angle=0, clip]{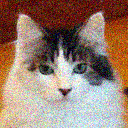} & 
    \includegraphics[width=0.140\textwidth, angle=0, clip]{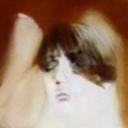} &
    \includegraphics[width=0.140\textwidth, angle=0, clip]{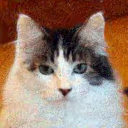} &
    \includegraphics[width=0.140\textwidth, angle=0, clip]{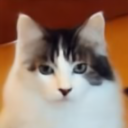} &
    \includegraphics[width=0.140\textwidth, angle=0, clip]{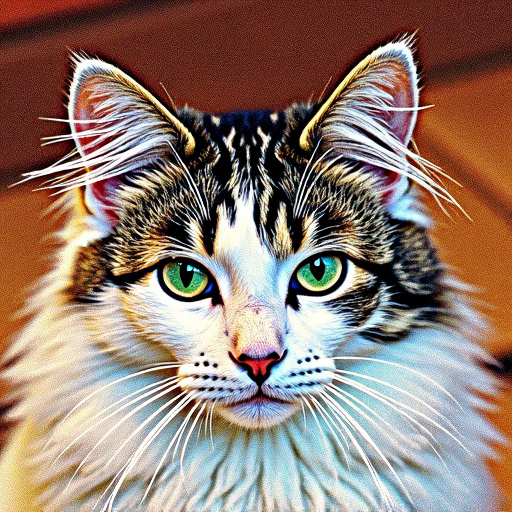} &
    \includegraphics[width=0.140\textwidth, angle=0, clip]{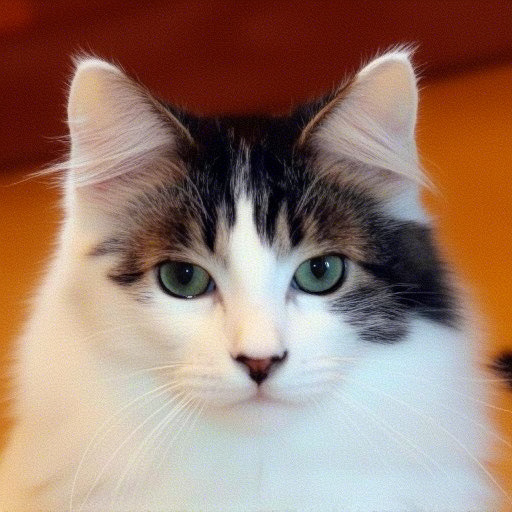} &
    \includegraphics[width=0.140\textwidth, angle=0, clip]{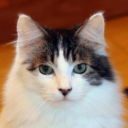}\\

    &
    \includegraphics[width=0.140\textwidth, angle=0, clip]{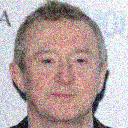} & 
    \includegraphics[width=0.140\textwidth, angle=0, clip]{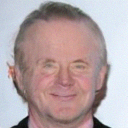} &
    \includegraphics[width=0.140\textwidth, angle=0, clip]{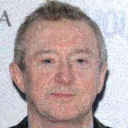} &
    \includegraphics[width=0.140\textwidth, angle=0, clip]{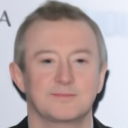} &
    \includegraphics[width=0.140\textwidth, angle=0, clip]{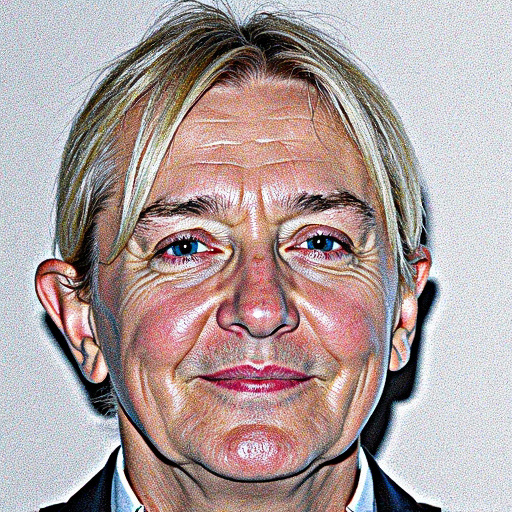} &
    \includegraphics[width=0.140\textwidth, angle=0, clip]{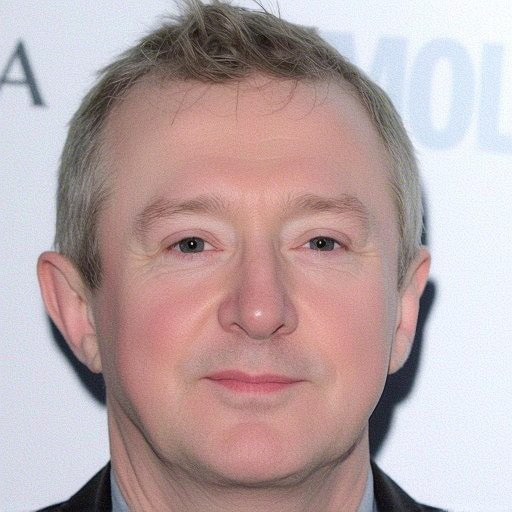} &
    \includegraphics[width=0.140\textwidth, angle=0, clip]{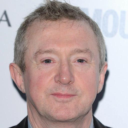}\\[0.1cm]

    \multirow{2}{*}[4ex]{\rotatebox{90}{\hspace*{3em}\textbf{Deblurring}}} &
    \includegraphics[width=0.140\textwidth, angle=0, clip]{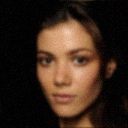} & 
    \includegraphics[width=0.140\textwidth, angle=0, clip]{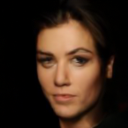} &
    \includegraphics[width=0.140\textwidth, angle=0, clip]{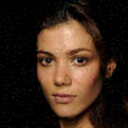} &
    \includegraphics[width=0.140\textwidth, angle=0, clip]{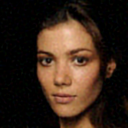} &
    \includegraphics[width=0.140\textwidth, angle=0, clip]{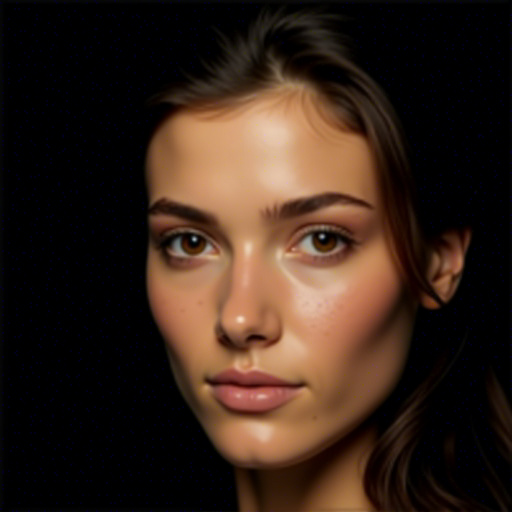} &
    \includegraphics[width=0.140\textwidth, angle=0, clip]{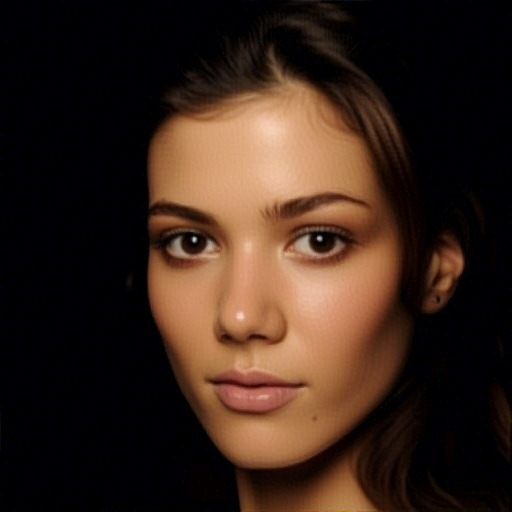} &
    \includegraphics[width=0.140\textwidth, angle=0, clip]{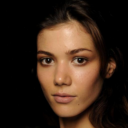}\\
    
    &
    \includegraphics[width=0.140\textwidth, angle=0, clip]{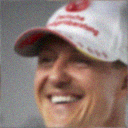} & 
    \includegraphics[width=0.140\textwidth, angle=0, clip]{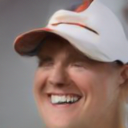} &
    \includegraphics[width=0.140\textwidth, angle=0, clip]{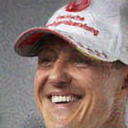} &
    \includegraphics[width=0.140\textwidth, angle=0, clip]{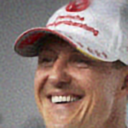} &
    \includegraphics[width=0.140\textwidth, angle=0, clip]{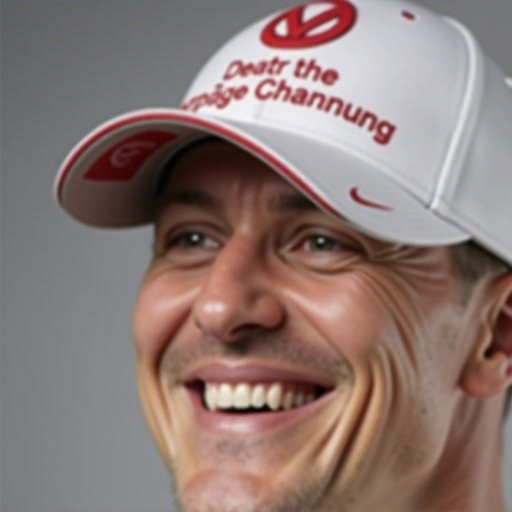} &
    \includegraphics[width=0.140\textwidth, angle=0, clip]{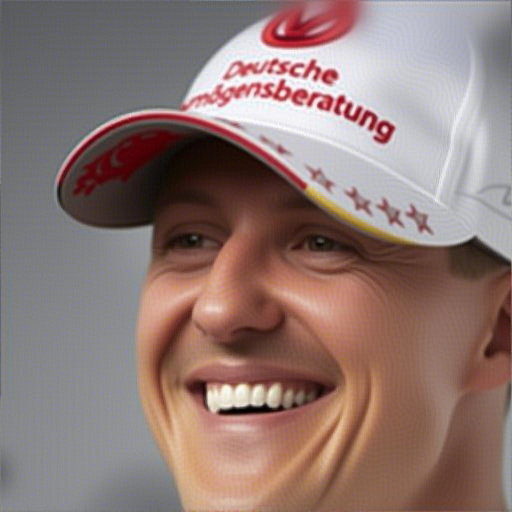} &
    \includegraphics[width=0.140\textwidth, angle=0, clip]{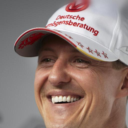}\\
    \rule{0pt}{4ex}  & Degraded  & D-Flow~\cite{ben2024_dflow} & {FlowPriors~\cite{zhang2024_flowpriors}} & PnP-Flow~\cite{martin2024pnpflow} & RFEdit~\cite{wang2024_RFEdit} & Ours & Clean \\
\end{tabular}
\endgroup
\caption{More qualitative comparisons of flow-based methods against FlowSteer. The restoration models in columns 2-4 have undesirable artifacts such as excessive blur. The image editing baseline in column 5 has poor fidelity. FlowSteer achieves better pixel-level fidelity, while generating viusally appealing details. Zoom in for better comparisons.}
\label{fig:qualitative_comparison}
\end{figure*}

\newpage
\FloatBarrier      
\clearpage         

\textbf{Acknowledgements:} We are thankful for Mykhailo Tsysin and Yu Yuan for the helpful discussions.
{
    \small
    \bibliographystyle{ieeenat_fullname}
    \bibliography{main}
}

\end{document}